# The Tychonic system: an implausible theory

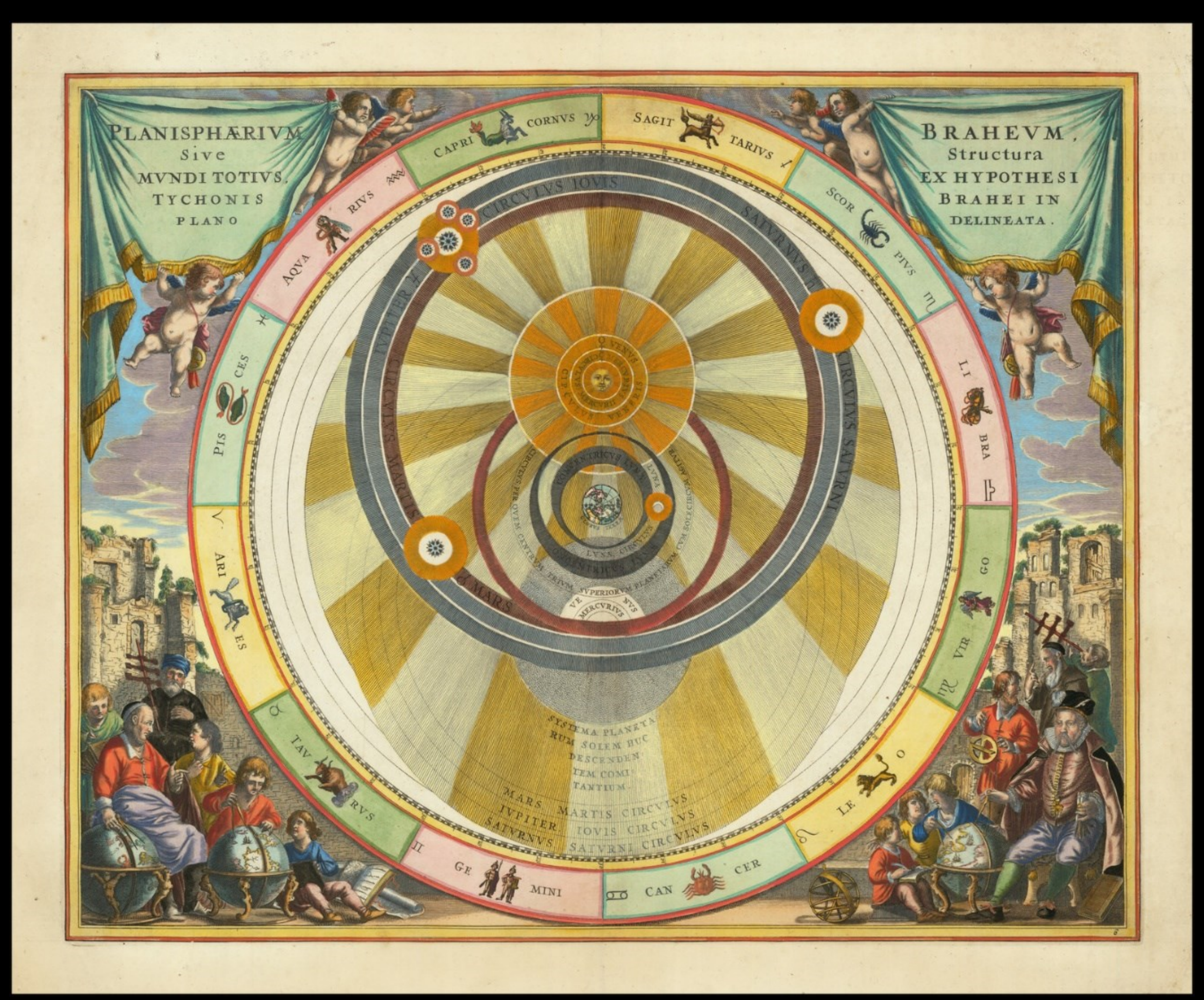

Gabriele Vanin

Gabriele Vanin

# THE TYCHONIC SYSTEM: AN IMPLAUSIBLE THEORY

For comments and communications: rheticus@tiscali.it

## 1. Galileo Galilei: not three systems of the world, but only two

Not a few historians have given a prominent place to the Tychonic system, believing it to be a credible alternative to the Ptolemaic and Copernican systems. On the contrary, several, perhaps the same ones, frowned upon Galileo's choice not to include it among the possible alternatives in his *Dialogo*, whilst he limited instead the discussion to Ptolemy and Copernicus. In fact, Galileo hardly mentioned it in his writings. If I am not mistaken, only two references to the Tychonic system can be found in his entire work. The first is found in Mario Guiducci's *Discorso delle comete*, which, as we know, is largely Galileo's work. While saying that in order to understand the true nature of comets, one would have to, echoing Seneca, know the entire structure of the universe, he states:

> ...therefore we must content ourselves with what the we can barely conjecture, however little, until the true constitution of the parts of the World is shown to us, for what was promised to us by Ticone remained imperfect.[1]

The second is in *Il Saggiatore*, where he downplays Brahe's cosmological contribution, as the latter did not fully develop it:

> ...I do not see for what reason Ticcone is elected, putting him before Tolomeo and Nicolò Copernico, whose two systems of the world we have in their entirety, and with great artifice constructed, and accomplished; which I do not see that Ticcone has done,...[2]

While, in the *Dialogo*, although he often cites Brahe's observations and praises him highly for his quality as an observer, he never once mentions his system.

## 2. Brahe reassessed?

Over the past few years, however, a number of contributions have appeared in favour of Brahe which, although starting from interesting points, have come to conclusions that seem somewhat far-fetched. It seems interesting to start my discussion precisely from these. In 2005, Harald Siebert,[3] of the Technische Universitat in Berlin and the University of Paris I, took up already known observations,[4] mentioned in notes and letters by Galileo and Benedetto Castelli,[5] of Mizar and the Trapezium. According to Siebert, they planned to use these stars for parallax measurements, on the assumption that the stars in a pair or group were randomly aligned along our line of sight and were at different distances from us (fig. 1). However, this is true for the former case, but there is no evidence for the latter in favour of this hypothesis, i.e. the observation may have been an end in itself (fig. 2).[6] Since Galileo measures the diameter of Mizar A at 6",[7] and makes the hypothesis that it is as large as the Sun, which has a diameter of 30', it should be at 300 AU and, at this distance, Siebert rightly says, it should show a parallax (the absolute one being 11.5') relative to Mizar B (always on the assumption that the latter is further away). This parallax should be easily measured with a telescope, which instead Galileo does not notice. Siebert rightly states that it is not known whether Galileo later made attempts in this direction. However, he claims that in the *Dialogo*, 15 years after these observations, Galileo attempted to prove that no one had ever tried to verify the Copernican theory with this experiment and to accredit himself as the only true Copernican in existence. But this is not true. In fact, in the *Dialogo* Galileo expressly says that the parallaxes, '"...if they were sought, were not [sought] in the proper manner, that is, with that accuracy, which to such minute punctuality would be needed;...[8]

[1] *Discorso delle comete di Mario Guiducci*, alle stelle Medicee, Florence, 1619, p. 44. Obviously, this consideration must be set within the scenario that once Ptolemy's fallacy had been demonstrated with a telescope, at the same time the edict of 1616 forbade support to Copernicus.
[2] *Il Saggiatore*, Mascardi, Rome, 1623, p. 25.
[3] H. Siebert, *The early search for stellar parallax: Galileo, Castelli, and Ramponi*, "Journal for the history of astronomy", 26, 3, 2005, pp. 251-271.
[4] Cf. U. Fedele, *Le prime osservazioni di stelle doppie*, "Coelum", 17, 949, pp. 65-69.
[5] *B. Castelli a [Galileo in Firenze]*, 16 novembre 1616 and *B. Castelli a Galileo in Firenze*, 7 gennaio 1617, in G. Galilei, "Opere", Barbera, Florence, 1890-1909, 12, pp. 296 and 301; Galilei, *Analecta astronomica*, "Opere", 3.2, pp. 877 e 880.

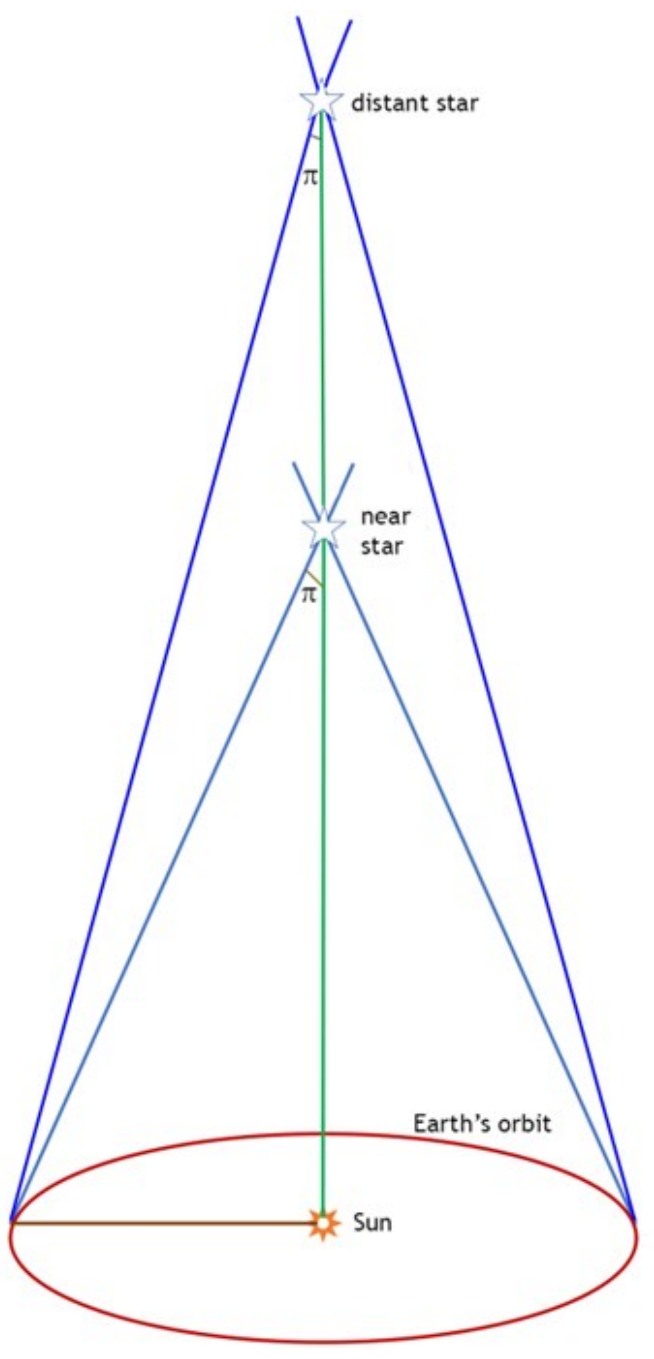

***Fig. 1.*** *A star that is closer than one that is further away has a greater parallax (denoted by π).*

Other questionable statements by Siebert are that until the end of the 18th century, the idea of multiple physical star systems seemed to be excluded a priori by heliocentrism and that Copernicans believed that all the stars in the universe were all exactly the size of our Sun, and that they were at the centre of other possible solar systems. Regarding the first point, this is not an a priori exclusion, but in fact, in the absence of the detection of orbital motions, this assumption could not be proven: rather than an a priori belief, it was an agnostic attitude. On the other hand, the fact that relative parallaxes could not be detected could have precisely this explanation, the existence of multiple systems. Not to mention that the presence of star clusters proved however that stars could be anything but isolated. Incidentally, throughout the 17th century there was very little attention paid to astronomical objects outside the solar system.[9] And well before the end of the 18th century, in 1767, John Michell,[10] on the basis of statistical considerations, stated that multiple star systems must have physical existence, well before Herschel announced that six of these systems had shown orbital variations.[11] Regarding the second point, Galileo himself recalls that it is possible to assume (and certainly not to affirm with certainty) that stars are not only equal in size[12] but also larger than the Sun.[13] It must also be remembered that Kepler, not just any Copernican, thought that the stars were much smaller than the Sun.[14] And that they were also surrounded by planets was an opinion that not all Copernicans held.

[6] Siebert reports the observation with a number of oversights:
First of all he says that if we followed Galileo's description we would not easily recognise the triple star he is talking about, but the description is in fact very clear;
that Galileo's star *g* is star TYC 4774-00953-1 of m 6.6, while it is star 4774-00931-1 of m 5.1;
that this star, together with star TYC 4774-00930-1, of m 6.3, which is missing in Galileo's drawing, forms a line parallel to Galileo's stars *a* and *b*, but this is not true: this star is star *c*, of m 6.7, in Galileo's drawing;
that it is not true that the line that passes through stars *c* and *i* is parallel to the one that passes through *a* and *b*, instead it is just so;
that Galileo states that stars *c*, *g* and *i* vary in brightness, whereas there is no trace of this statement;
that it is untrue that *i* and *c* are much fainter than *g* (four or five times says Galileo), whereas the statement is astonishingly correct, especially if we take into account the performance of a Galilean telescope, and that the mathematical progression corresponding to the various magnitude values was not yet known: *c* and *i* are 4.4 times fainter than *g*;
that the distances between them of *g*, *c* and *i* reported by Galileo are too small with respect to the stars of the Trapezium, whereas they are reported surprisingly exactly, 0.26 times the distance between *a* and *b*, as is the case in reality, an incredible figure if we take into account the lack of micrometric measuring devices.
Therefore Siebert believes that Galileo's triple star is formed by TYC 4774-00953-1, TYC 4778-01358-1 of m 7.2 which, Siebert says, is too bright to match Galileo's description but which, at the time, being a variable, may have been fainter, and HIP 26220 of m 11.1; however, it has already been pointed out that first identification is incorrect. Moreover, HIP 26220 coincides with TYC 4774-00953-1, and TYC 4778-01358-1 simply does not exist; not to mention that a star of m 11.1 is far beyond the reach of Galileo's telescope.
Furthermore, Siebert states that on the line between Galileo's stars *b* and *g* lies another star, TYC 4774-00931-1, which Galileo does not report. Instead the star mentioned is precisely Galileo's *g*.
It would have been enough for Siebert to have personally observed the Trapezium with a Galilean telescope or a small modern refractor with a 5 cm aperture, to realise how extremely accurate Galileo's observation was: by means of this type of instrument only three of the four stars of Trapezium can be seen, and they are precisely those described by Galileo, with very precise magnitudes and distances; the fourth one, TYC 4774-00931-1 of m 7.9, is in theory within his reach, but its proximity to TYC 4774-953-1 makes it in fact not perceptible: cf. Gabriele Vanin. *Much better than thought: observing with Galileo's telescopes*, "Journal for the history of astronomy", 46, 4, 441-468 (2015), p. 462.
[7] Galilei, *Opere*, 3.2, p. 877 (Siebert, however, on p. 259 misinterprets the half-diameters of Mizar A and B given by Galileo, 3" and 2" respectively, as the major and minor diameters of Mizar A).
[8] *Dialogo di Galileo Galilei Linceo*, Landini, Florence, 1632, p. 380.

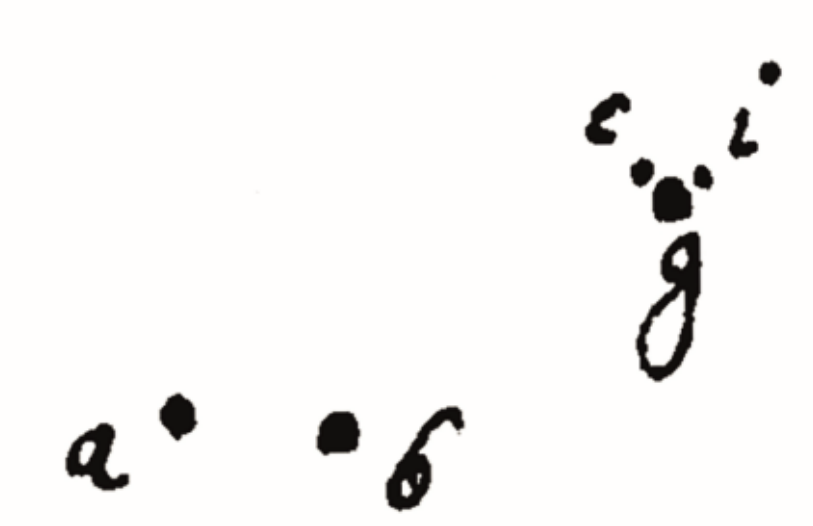

***Fig. 2.*** *The region of Orion's Trapezium drawn by Galileo in the* Analecta astronomica *(from Galilei,* Opere*, 3.2, p. 880).*

In 2007, in a largely acceptable article dedicated to the accuracy of Galileo's observations, Christopher M. Graney, of Jefferson Community College in Louisville, Kentucky,[15] argued that in the famous passage of the third day of the *Dialogo*, Galileo stated that stars of the first magnitude have a diameter of 5”, and those of the sixth magnitude 5/6”, whereas, clearly,[16] these are only upper limits, not absolute ones. This error would be repeated by Graney in all the numerous subsequent articles he would devote to Galileo's stellar observations.

In 2008, Graney, echoing Siebert's work, pointed out that Galileo measured stellar diameters in the belief that they were real,[17] as he thought that there was a linear relationship between apparent dimensions and magnitudes, and hence distances.[18]

In another article in 2008,[19] after pointing out what Galileo says in the *Dialogo*[20] about the fact that in the future it may be possible to evidence a parallel shift in the Earth's motion, he does state that Mizar had already testified, thus proving Galileo wrong, and that he could therefore have supported, instead of the heliocentric theory, Tycho Brahe’s theory , which Robert Hooke, still in 1670, considered a viable alternative. The last statement is a bit of a stretch, since Hooke was an ardent Copernican, one of the first advocates of the principle of inertia, of the complete identity between terrestrial and celestial motions, of the force of

---

[9] This did not prevent the Jesuit Athanasius Kircher, as Siebert also reports, from imagining stars orbiting each other in a work of fiction, although he did so from an anti-Copernican perspective: cf. *Itinerarium extaticum*, Mascardi, Rome, 1656.

[10] J. Michell, *An inquiry into the probable parallax, and magnitude of the fixed stars, from the quantity of light which they afford us, and the particular circumstances of their situation*, “Philosophical Transactions”, 57, 1767, pp. 243-250.

[11] W. Herschel, *Account of the changes that have happened, during the last twenty-five years, in the relative situation of double-stars; with an investigation of the cause to which they are owing*, “Philosophical Transactions”, 93, 1803, pp. 339-382.

[12] Galilei, *Opere*, 3.2, pp. 876, 877 e 878.

[13] G. Galilei, *Lettera a Francesco Ingoli* (1624), in *Opere*, 6, pp. 524 e 525.

[14] J. Kepplero, *Epitome in astronomiae copernicanae*, Plancus, Linz, 1618-21, p. 498.

[15] C. M. Graney, *On the accuracy of Galileo’s observations*, “Baltic astronomy”, 16, 2007, p. 446.

[16] Cf. Galilei, *Dialogo*, p. 352; a notion also found in *Lettera a Ingoli*, p. 533. In the *Dialogo*, he also says that Sirius’ diameter is no more than 3” (“the irradiated is shown to be more than a thousand times greater than the naked”, p. 69) and no more than 2” (“it will not be judged to be the twentieth part of that of Jupiter”, p. 329). In his letter of February 1611 to Giuliano de' Medici (*Opere*, 11, p. 62) he had stated that the diameter of Sirius was “not greater than the fiftieth part of that of Jupiter”.

[17] C. M. Graney, *Objects in telescope are farther than they appear*, “ArXiv”, 2008, pp. 2-3.

[18] *Op. cit*, p. 5.

[19] C. M. Graney, *But Still, It Moves: Tides, Stellar Parallax, and Galileo’s Commitment to the Copernican Theory*, “Physics in perspective”, 10, 3, 2008, pp. 258-268.

[20] Galilei, *Dialogo*, p. 375.

gravity and its dependence on the inverse of the square of distance. In this regard, Graney mentions Hooke's writing in which he announces having found the parallax of Eltanin, in Draco (fig. 3).[21] Thus, if anything, he mentions having realised the experience with which he wanted to discriminate between the Copernican theory and the Tychonic one (given that Ptolemy was now out of the picture thanks to Galilean observations). In this writing Hooke expresses himself harshly towards those who, "either out of ignorance or prejudice", did not adopt Copernicus, while "the more knowing and judicious" did.[22] In fact, Hooke's judgement on the Tychonic system is not very different from that on Ptolemy, when he writes:[23]

> Now though it may be said, 'tis not only those but great geometrician, astronomers and philosophers have also adhered to that side, yet generally the reason is the very same. For most of those, when young, have been imbued with principles as gross and rude as those of the vulgar, especially as to the frame and fabrick of the world, which leave to deep an impression upon the fancy, that they are not without great pain and trouble obliterated: others, as a further confirmation in their childish opinion, have been instructed in the *Ptolomaick* or *Tichonick* system, and by the authority of their tutors, over-awed into a belief, if not a veneration thereof: whence for the most part such persons will not indure to hear arguments against it, and if they do, 'tis only to find answers to confute them.

Graney also says that when, before the Holy Office, Galileo claimed that the purpose of the *Dialogo* was to refute Copernicus, not to support him, it would have been much more convincing if he had published his negative observations on double stars at the time - an irony, if irony at all, that is frankly misplaced and slightly macabre. The conclusions of the article are puzzling, as Graney states that Galileo found the heliocentric theory so convincing that he supported it even when observations of tides and parallaxes did not agree with his predictions. As if lack of evidence automatically meant contradiction, and as if Galileo had systematically researched stellar parallax using double stars, which that does not emerge from his books, letters or manuscripts. Even stranger is the assertion that if Galileo had published his observations on Mizar, the acceptance of Copernicus' theory would have been significantly delayed: in the first place, Galileo says in the *Dialogo* that parallax has not yet been found, and this would not have changed if he had published the observation. Secondly, it was a single star, so what probative force could this single observation have? Thirdly, there were quite other reasons, as is well known, to support Copernicus. Fourthly, Galileo himself, in the *Dialogo*, gives higher (minimum, we always remember) stellar distances,[24] and other Copernicans thought that the stars were much farther apart than Galileo estimated in the *analecta*. Copernicus himself believed, as Galileo remembers,[25] that the sphere of the fixed was at an immense distance and Kepler established that it was at a distance 58 times greater than that given in the *analecta*.[26]

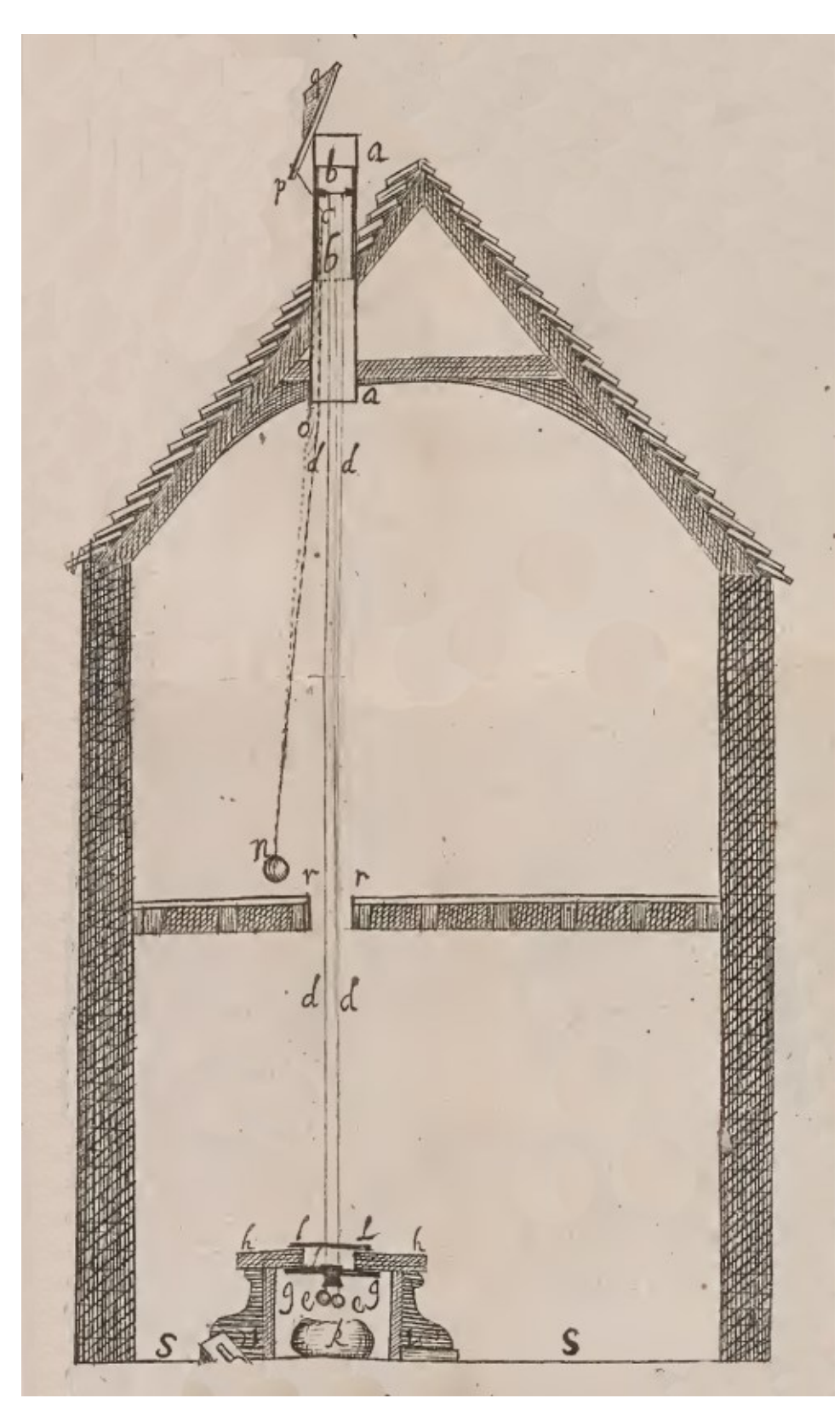

***Fig. 3.*** *The zenith telescope with which Hooke intended to measure Eltanin's parallax (from Hooke,* An attempt to prove the motion of the Earth*).*

[21] R. Hooke, *An attempt to prove the motion of the Earth from observations*, in *Lectiones Cutlerianae*, Martyn, London, 1679, pp. 1-28. In fact, Hooke probably anticipated, like Rømer and Flamsteed, the discovery of the aberration of light (see R. Grant, *History of physical astronomy*, Baldwin, London, 1852, pp. 548-549).
[22] Hooke, *op. cit*., p. 1.
[23] *Op. cit*., p. 3.
[24] Galilei, *Dialogo*, p. 352.
[25] *Op. cit*., p. 380.
[26] Kepplero, *Epitome*, p. 497.

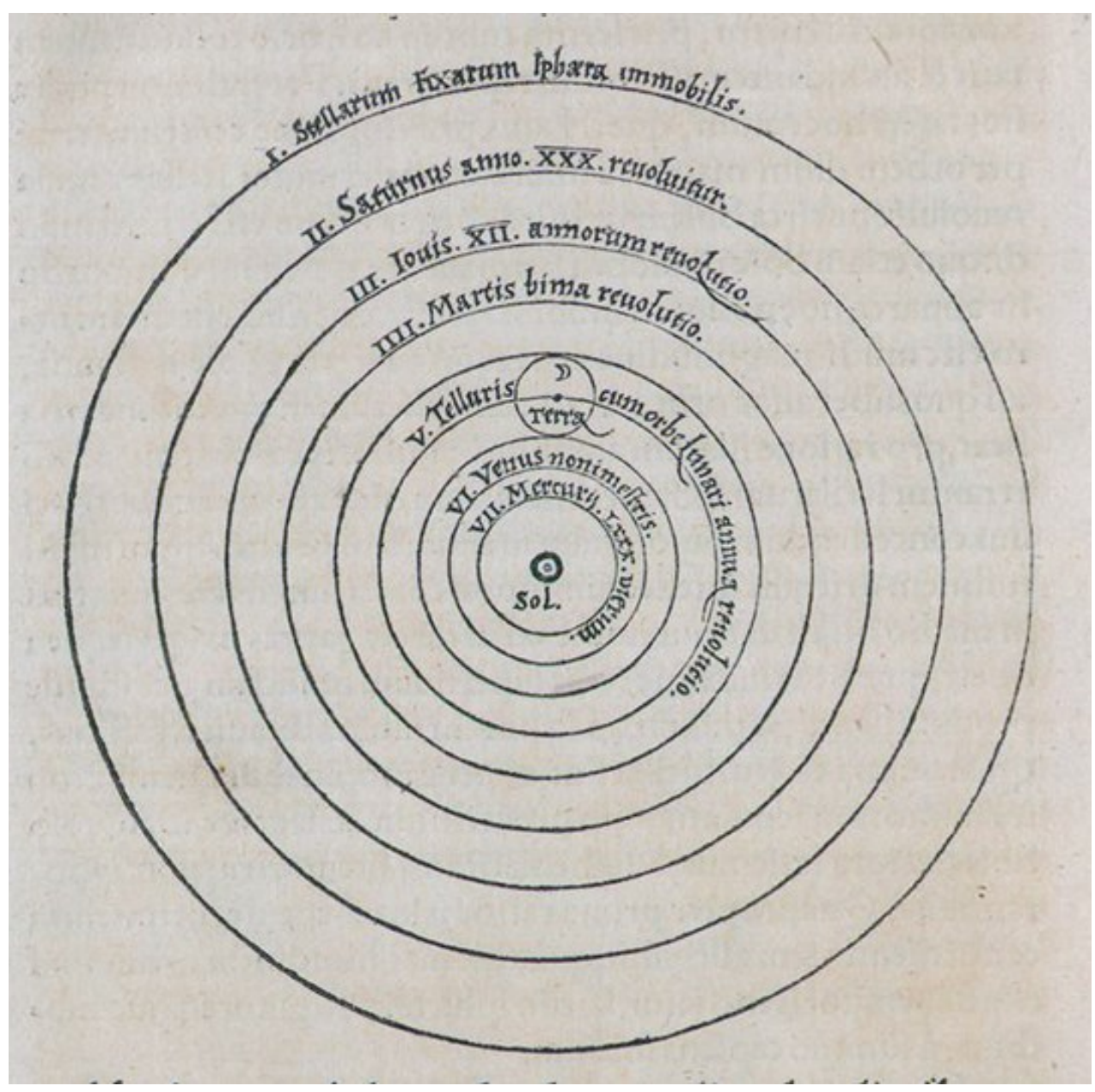

*Fig. 4. The Copernican system (from* De revolutionibus orbium coelestium*).*

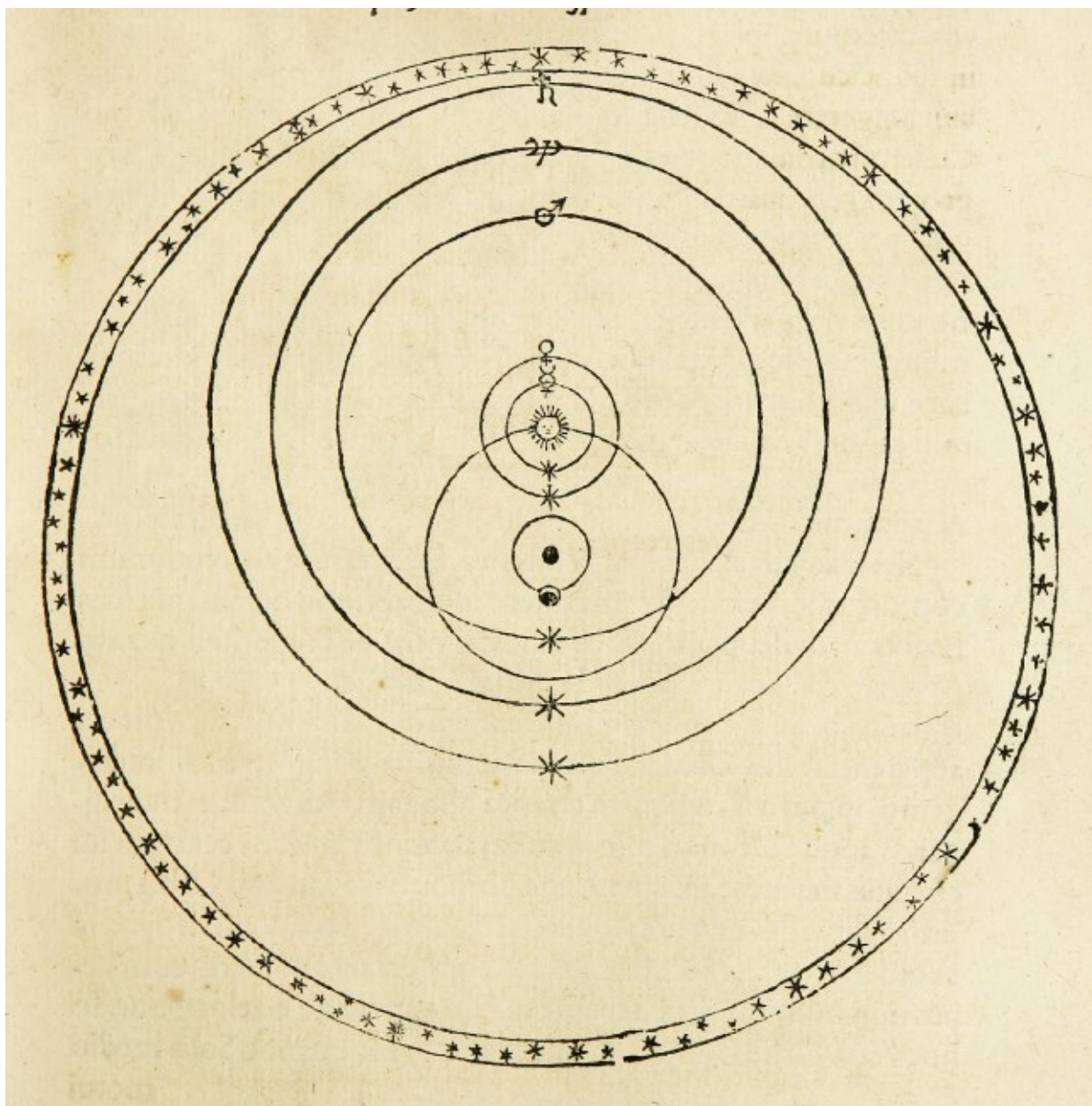

*Fig. 5. The Tychonic system (from* De mundi aetherei recentioribus phaenomenis*).*

In a different article in 2008, Graney goes so far as to say, in a completely anachronistic manner, that Galileo promoted the so-called "Copernican principle"[27] (more appropriately, the "cosmological principle"), i.e. a concept introduced by Einstein with general relativity to indicate a universe that looks the same no matter where it is observed from.[28] He presents a mental experiment of star-counting that Galileo could have done and which might have proved him right,[29] but which appears frankly quite inapplicable with the methods and procedures used at the time. Even reducing the matter to the bone, however, i.e. to the fact that the number of stars increases with increasing magnitude, something that could already be seen with Ptolemy's catalogue (at least up to the fourth magnitude), it is not true that there is no reason why this should be so if the stars are attached to a solid sphere, as Graney claims. In fact they could all be at the same distance but have different intrinsic brightness, with the brightest ones being fewer in number than the weakest ones, so much so that Kepler, for example, still believed in the existence of the starry sphere. And so it is certainly not proof, as Graney claims, that the stars are many suns distributed in space, as Galileo believed. So one does not see the point of introducing such a theoretical experiment. Finally, Graney uses an obviously different weight in evaluating the work of Galileo and Tycho. For while in the previous articles he argued that Galileo, not measuring a stellar parallax, might have converted to the Tychonic system, here he claims that Tycho, not finding the parallax of Mars to settle the dispute between the heliocentric and geocentric models, resolved to develop his own model.[30] But one cannot see the reason for this. Since Brahe admitted a parallax of 3' for the Sun, both in the Copernican model (fig. 4) and in his own (fig. 5), the planet, which comes much closer to the Earth than the Sun, should have presented a parallax between 4.5' and 8' in opposition, within the reach of even the most meagre of his instruments. Then, as he had not found it, Tycho should have remained Ptolemaic.

[27] C. M. Graney, *Visible stars as apparent observational evidence in favor of the copernican principle in the early 17th century*, "Baltic astronomy", 17, 2008, p. 426.
[28] H. Bondi, W.H. Bonnor, R.A. Lyttleton, G.J. Whitrow, *Rival theories of cosmology*; I am quoting from the Italian edition: *Teorie cosmologiche rivali, Einaudi, Turin, 1965, pp. 16 and 50.*
[29] Graney, 2008, o*p. cit*., pp. 432-437.
[30] *Op. cit.*, pp. 431-432.

Or, alternatively, since his measurements were accurate to within 1', he should have adopted a parallax of 30" for the Sun, much closer to reality.
In 2009, Graney and Henry Sipes, of Samtec Inc., published an article very similar to the previous one but in it they wrote that Galileo had observed a "certain number" of multiple stars, whereas in fact we have no knowledge of any other observations apart from those of Mizar and Trapezium.[31] They too claim[32] that Galileo believed that the stars were all as big as the Sun, whereas, as we have seen, this was only a working hypothesis referring to a specific case, which, moreover, was never published. Here again, a mathematical model of double stars that Galileo might have observed is proposed, but it is not seen as necessary, even for the purposes of the conclusions reached in the article. They also state[33] that since Galileo finds no relative parallax between the stars of the Trapezium, one of the possible hypotheses is that they constitute a physically bound system; but since Galileo states that stars *c* and *i* have apparent diameters of only a quarter or a fifth of *g*,[34] their physical diameters must actually vary by a factor of four or five, and that such a variation would be a fatal break from Galileo's thoughts, magnified by the possibility of Galileo's observation of other pairs with stars of different magnitudes. But it is hard to see why. First of all, as mentioned above, there is no evidence that the observation of the Trapezium was aimed at finding parallaxes. Secondly, Galileo speaks of *magnitudo* referring, as we have already seen, not to the apparent size of the stars, but to their luminosity (see footnote 6). Hence the difference in apparent size is smaller and, according to the diameters shown in the drawing (see Fig. 2), less than a factor of 2, consistent with Galileo's assertion on stellar dimensions (see footnotes 12, 13 and 16), and thus not excluding that they could be found at the same distance from us. Truly noteworthy is the close of the article, where it is stated that seventeenth-century astronomers would have come to the conclusion that Galileo's ideas on the structure of the universe were wrong, because it does not consist of Sun-like stars distributed in space. A more persuasive interpretation of the data would have been one that supported Copernicus' original description, with the sphere of the fixed centred on the Sun, and an even more persuasive one, explaining the data in the best possible way, i.e. the Tychonic one. Not those assuming that the stars did not show parallax because they were further apart or because they were physically linked!
In 2010, Graney alone went back to the issue, summarising what had been said in the two previous articles.[35] While he had previously used the expedient of resorting to conceptual experiments that no seventeenth-century astronomer had performed nor could have performed, this time, we do not know why, he simulates the existence of a fictitious astronomer and attributes to him observations that were actually made by two real astronomers, Galileo and Mayr. In the introduction, Graney states that the observation of the stars is the only approach that can determine which of the two systems, the Copernican and the Tychonic, is valid. The statement, as we will say later, appears to be meaningless. He then states that a telescope such as Galileo's one showed an Airy disk of 10", while the correct value (which can be deduced from the formula he himself gives in a note) is 3". Even more strangely, he states that Galileo's hypothesis that the stars are suns distributed in space, and not all at the same distance attached to a solid sphere, supports the view that the Earth moves around itself and not that the stars revolve around the Earth, but that the impossibility of revealing the relative stellar parallax in double stars shows that the Earth does not revolve around the Sun, but it is the latter that revolves around the Earth. This is congruent not with the Copernican system, but with the Tychonic one. He seems to forget, however, that Tycho did not even accept the rotation of the Earth on its axis. Indeed, Tycho's problem was that he just could not accept that a body as heavy as the Earth could move: if he had accepted that it rotated on its own axis, it is hard to see why he could not also accept its revolution. Besides, Graney comments on the observations of Simon Mayr, who, like Galileo, can see round stars with a telescope but argues, unlike Galileo, that this fact supports the Tychonic view. Graney rightly considers that since Mayr does not provide any details to justify this hypothesis, this observation alone is not convincing at all, because the stars could still be very large bodies placed at various large distances, or very large lights of various sizes attached to the celestial sphere. Nonetheless, since Mayr states that by being round, the stars cannot be as distant as Copernicus claims, Graney contradictorily tends to assume that this view is more consistent with observational data than Galileo's. One should also remember that Mayr does not provide any angular measurements of stars, nor does he assume any physical dimensions for them, nor does he try to measure relative parallaxes of double stars. If he had, for example, found that the stars had diameters of 3”, an intermediate value among all those found by Galileo, and thought that the stars were as large as the Earth, they would be over 100 times farther away than the Sun. More, had he thought that they were as large as the Sun, they would have been 600 times farther away than our star. Wouldn't that be enough to prove Copernicus right? Even taking into account the fact that Tycho was unable to detect any stellar parallax with his instruments (assuming Mayr was aware of this), within then 1', to account for this deficiency it would be enough to assume stellar diameters five times larger than the Sun. Not an unreasonable assumption, particularly if one considers that at that time the Sun was only assigned a diameter of 1/20th of its real one. Graney also states that the Tychonic system exerted a strong appeal in the 17th and 18th centuries, as seen in Giovanni Battista Riccioli's *Almagestum Novum* of 1651 and Johann Gabriel Doppelmayr's *Atlas Coelestis* of 1742. However, Riccioli proposed a system (similar-Tychonic, actually, not Tychonic) for doctrinal reasons, as we shall see, whereas Doppelmayr was not a leading astronomer, but a populariser, and in his work he paid attention to the Tychonic system for historical reasons, as is still done in popular books today.

---

[31] C. M. Graney and H. Sipes, *Regarding the potential impact of double star observations on conceptions of the universe*, "Baltic Astronomy", 18, 2009, p. 95.
[32] *Op. cit.*, p. 105.
[33] Ibid.
[34] Galilei, *Opere*, 3.2, p. 880.
[35] C. M. Graney, *Seeds of a Tychonic revolution: telescopic observations of the stars by Galileo Galilei and Simon Marius*, "Physics in perspective", 12, 1, 2010, pp. 4-24.

In 2010, Graney[36] also presents Riccioli's work on stellar dimensions.[37] Riccioli states that the measurement of the apparent diameters of stars made with a telescope proves Tycho right and Copernicus wrong. This is not only because of the failure to detect stellar parallax, but also because, following Tycho (for whom the apparent diameters of stars ranged from 20” for the faintest to 2'15” for the brightest),[38] one would have to admit that stars’ sizes are enormous, comparable to the size of the Earth's entire orbit. Riccioli criticises Galilei on the method of his measurements as reported in the *Dialogo*, which he defines misleading. However, the measurements he publishes, comparing the stellar discs to the dimensions of Jupiter and Saturn, seem truly excessive: 18” for Sirius against Galileo's 5”,[39] 7” to 12” for second magnitude stars against Galileo's 5”-6”,[40] 7” for third magnitude stars against Galileo's 4”,[41] 6” for fourth magnitude stars against Galileo's 4”.[42] It is in fact very strange that he, with better telescopes than Galileo's, reports values on average double those measured by Galileo (with the telescope, not with the method mentioned above). However, using Riccioli's data on apparent diameters,[43] the actual diameters of stars with a parallax of less than 1' could not be less than 0.037 times the diameter of the Earth's orbit for the faintest stars and 0.15 times for the brightest ones. However, since Riccioli believed that the advent of the telescope had lowered the limit of stellar parallax detection to 10”, the diameters were between 0.22 and 0.9 Earth orbital diameters, i.e. back to Tycho's values. Between 2011 and 2013, Graney returned to the subject by devoting three articles, very similar to each other, to exposing Riccioli's crucial objection to Copernican astronomy. The conclusion of the first article is noteworthy:[44]

---

[36] C. M. Graney, *The telescope against Copernicus: star observations by Riccioli supporting a geocentric universe*, “Journal for the history of astronomy”, 41, 4, 2010, pp. 453-467.

[37] G. B. Riccioli, *Almagestum novum*, Benati, Bologna, 1651, 1, pp. 715-717.

[38] *Tychonis Brahe dani, Astronomiae instauratae progymnasmata*, Tampachius, Frankfurt, 1610, p. 482.

[39] Galilei, *Opere*, 3.2, p. 878.

[40] *Op. cit.,* pp. 876 e 877.

[41] *Op. cit*., p. 877.

[42] Ibid.

[43] The tables of stellar distances obtained for various parallaxes and the resulting stellar diameters he published in *Almagestum Novum* are unusable because they are full of errors.

[44] C. M. Graney, *Regarding how Tycho Brahe noted the absurdity of the copernican theory regarding the bigness of stars, while the copernicans appealed to God to answer that absurdity*, “ArXiv”, 2011, p. 9.

> We say that Tycho's opposition to the Copernican theory was related to religion or the world view of his time or just his inability to take the next logical step. We do not even think of the bigness of stars, and how that bigness once rendered the heliocentric theory so irrational, inharmonious, and disproportionate that Copernicans could only appeal to the power of God to explain it all.

In the second article,[45] he refers to Riccioli's assertion that the only evidence Copernicans object to the argument of enormous stellar dimensions is the power of God. In fact, the Copernican Christoph Rothmann, responding to Tycho on this, had stated that:[46]

> What is absurd about a star of the third magnitude equalling the entire annual orbit? What is there in this that opposes the divine will, or is impossible for divine nature, or inadmissible for infinite nature? You must demonstrate all this thoroughly if you wish to deduce anything absurd from it. What the vulgar sees at first sight as absurd we cannot easily regard as such, for divine wisdom and majesty are far greater. And however much we concede the vastness and size of the universe, they will still not be proportionate to the infinite Creator. The greater the king, the greater the palace befitting his majesty.

However, Riccioli's statement, to which Graney gives credit, does not seem correct. In fact, it is not, strictly speaking, a theological explanation, as much as a kind of appeal to common sense, particularly if we replace the word "God" with the word "nature", as Rothmann does at one point: in other words, Rothmann says, what is there, in fact, that prevents the stars from being as big or small as desired? Besides, if we think of God as an all-powerful being with immense powers, why not think that the universe he has created is, in his image and likeness, much bigger than we have hitherto imagined?
Graney, on the other hand, intensifies his view, stating that the same ideas were held by Lansbergen, Digges and Copernicus himself. Lansbergen actually echoes Rothmann's claim that the sphere of the fixed is of almost infinite size,[47] and that every star in it is as large as the Earth's orbit.[48] However, it must be pointed out that Lansbergen was a fervent Calvinist priest and preacher, and that his reasoning, with no great pretensions to scientificity, was contained within a popularising book of Copernican ideas, not in a scientific treatise. Digges, on the other hand, refers generically to the universe as an infinite place, consistent with a god of infinite power although he never mentions stellar dimensions.[49] Finally, the quotation from Copernicus is unrelated, since he merely says: "How extremely fine is the divine work of the best and greatest artist!"[50] It must also be said that at that time, references to God were common in scientific works, since everyone reasoned in theological terms, theology being the most important discipline and the reference substratum of all others. Nonetheless, Graney writes: "Riccioli may have assumed that an appeal to the power of God regarding the fixed stars was implied in the Copernican hypothesis." Indeed!
At the end of the article Graney provides a defence of the Tychonic hypothesis which appears perhaps a little too simplistic, that is:

> … has its own awkward points, particularly in regards to how the Sun tows the planets around with it, but the explanations (such as a fluid planetary heaven) needed to deal with these problems are relatively straightforward. In the Tychonic hypothesis the sphere of the elements is at rest, heavy bodies have one natural rectilinear motion, stars are reasonable in size, projectiles and falling bodies aren't expected to exhibit odd deflections that do not appear, and God isn't invoked to make it all work.

---

[45] C. M. Graney, *Science rather than God: Riccioli's review of the case for and against the copernican hypothesis*, "Journal for the history of astronomy", 43, 2, 2012, pp. 215-225.
[46] *Letter of 18 April 1590 from Rothmann to Brahe*, in *Tychonis Brahe Dani epistolarum astronomicarum libri*, Uraniborg, 1596, p. 186.
[47] *Philippi Lansbergii commentationes in motum Terrae*, Romanus, Middelburg, 1630, p. 52.
[48] *Op. cit.*, p. 51.
[49] T. Digges, *A perfit description of the celestiall orbes*…, in L. Digges, *A prognostication everlasting of right good*…, Kingston, London, 1605, ff. N1v, N2r.
[50] *De revolutionibus orbium coelestium*, Petreius, Nuremberg, 1543, p. 10r.

In the third article,[51] Graney makes the same point as in the previous one, but expounds more extensively on Lansbergen's thought, from which it emerges quite clearly that his work, rather than a treatise on astronomy, was a theological exposition, with many biblical quotations. Somehow over-the-top is Lansbergen's metaphor that the stars are the soldiers, God's army, through which he achieves everything he wants.[52] It is then somewhat perplexing that Graney suggests the idea that these views might have been representative of the Copernicans in general at the time.[53]

## 3. The genesis of Brahe's system

As well known, a mixed planetary theory, with the inner planets revolving around the Sun, had been already proposed in antiquity (fig. 6).[54] In 1500, an Indian astronomer, Nilakantha Somayaji, extended the model to the higher planets.[55] Prior to 1553 (the date of his death) Erasmus Reinhold, as revealed by a letter from Rothmann to Tycho[56] and by the discovery of an unpublished manuscript by Reinhold himself by Alexander Birkenmajer,[57] had devised a system in which he performed a purely mathematical "inversion" of the Copernican system, in which all eccentricities, epicyclic dimensions, and distances found by Copernicus were retained, but the Sun was thought to be responsible for the motion of the Earth.

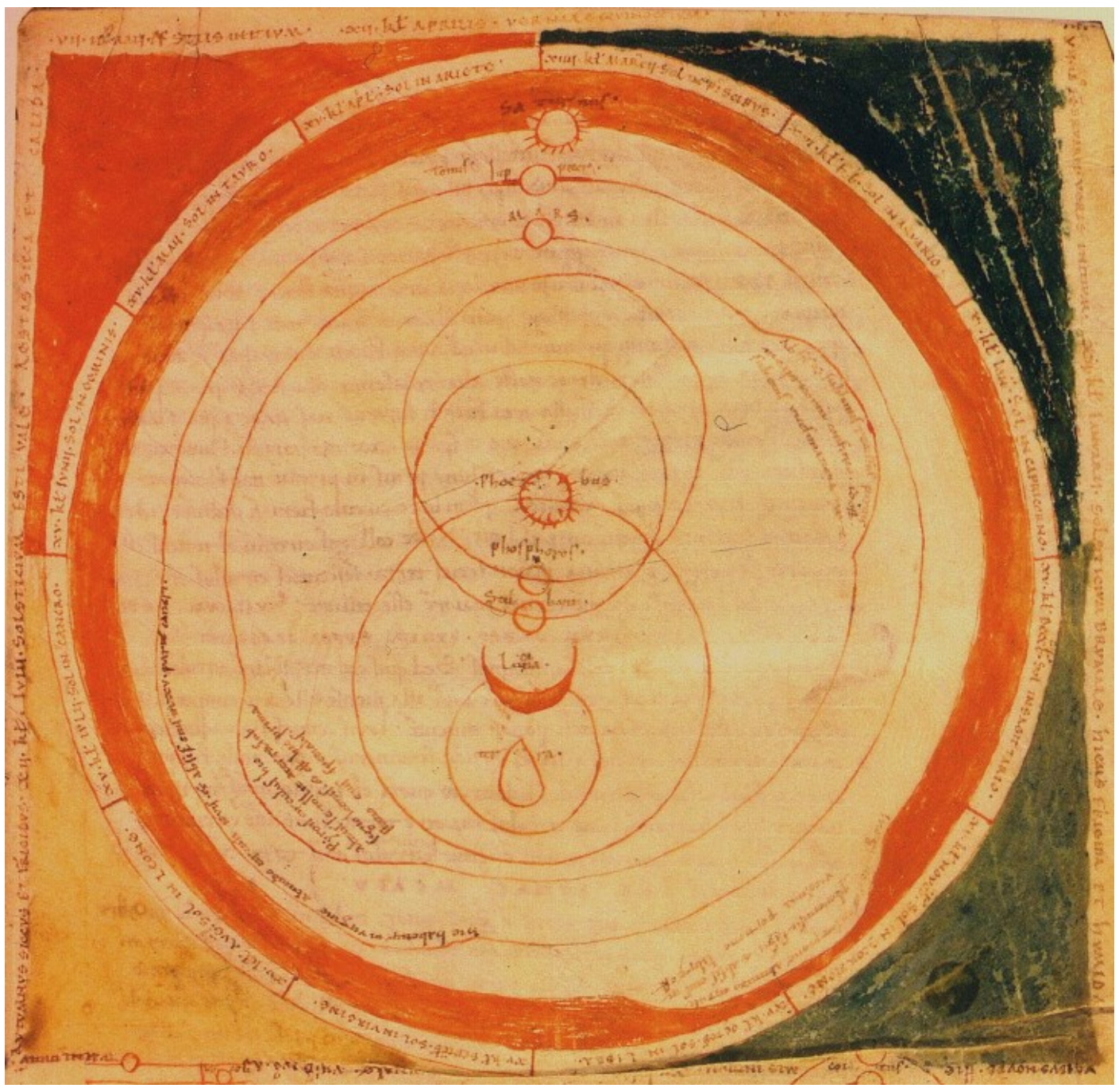

***Fig. 6.*** *The "mixed" astronomical system, in which Mercury and Venus revolve around the Sun, represented here in a manuscript of* De nuptiis Philologiae et Mercurii *by Marziano Capella, 11th century, Firenze, Biblioteca Medicea Laurenziana, San Marco 190, fol. 102r.*

In 1573, the German astronomer Valentin Naboth, in a textbook for gymnasium students, reproduced a diagram representing the ancient mixed theory limited to the inner planets (fig. 7).[58] In 1574, Tycho Brahe, in a lecture of a course he was to give at the University of Copenhagen (which only ended after a few lectures due to his transfer to Germany), expressed his intention to teach students an inverted version of the Copernican system, one with the Earth at rest.[59] Probably influenced by Naboth, another German astronomer, Paul Wittich, adopted the mixed system for Mercury and Venus in 1578 in one of his annotated copies of *De revolutionibus* (fig. 8) as one of the working hypotheses to explain the motions of the two planets.[60] Between 1577 and 1578, Tycho Brahe adopted the same model.[61] In the summer of 1580, Wittich spent three months in Uraniborg as Brahe's assistant and showed and discussed his cosmological diagrams with him.[62] In 1583 Brahe thought of extending the model to the outer planets as well.[63]

[51] C. M. Graney, *Stars as the armies of god: Lansbergen's incorporation of Tycho Brahe's star-size argument into the copernican theory*, "Journal for the history of astronomy", 44, 2, 2013, pp. 165-172.
[52] *Lansbergii, Op. cit*., p. 58.
[53] An article summarising Graney's main arguments was also published: D. Danielson and C. M. Graney, *The case against Copernicus*, "Scientific American", 310, 1, 2014, pp. 72-77.
[54] See G. Vanin: *Sulla rimozione della teoria eliocentrica nell'antichità*, "Giornale di Astronomia", 48, 2, 2022, pp. 24-25.
[55] K. Ramasubramanian, *Model of planetary motion in the works of Kerala astronomers*, "Bulletin of the Astronomical Society of India", 26, 1998, pp. 11-31.
[56] *Letter of 1 October 1588 from Rothmann to Brahe*, in *Tychonis Brahe Dani epistolarum astronomicarum libri*, p. 127.
[57] C. J. Schofield, *Tychonic and semi-Tychonic world systems*, Arno Press, New York, 1981, p. 23.
[58] *Primarum de coelo et terra institutionum quotidianarumque mundi revolutionum,* Venice, 1573, p. 41v.
[59] T. Brahe, *Opera omnia* (ed. by J.L.E. Dreyer), Hauniae In Libraria Gyldendaliana, Copenaghen, 1913-29, 1, p. 172.

Around the same time, three German astronomers came up with the same results: the first was in 1585 Nicolaus Reimers, known as Ursus,[64] who was accused by Tycho of plagiarism, though historical criticism has not been able to prove this irrefutably. However, he suggested that the Earth rotated on its own axis (a system which is also called, perhaps improperly, semi-Tichonic) and the orbit of Mars did not intersect that of the Sun, being completely outside the latter's orbit (fig. 9). Second came, between 1586 and 1587, Rothmann himself, who was probably using this as an expedient to make Copernicus' theory more acceptable by scholars.[65] The third astronomer, in 1588, Helisaeus Roeslin, was similarly accused by Tycho of plagiarism, but declared in a letter to Maestlin that he was unaware of Tycho's system and had learned of the existence of his own from Ursus. However, both of these systems did not convince him because they were, in his opinion, incapable of explaining how the planets could follow the Sun in its annual orbit around the Earth, in the absence of solid spheres conducting the motion.[66] In a book published in 1597,[67] he presents and comments on the Ptolemaic system, the Copernican system, Ursus' system, Brahe's system and his own, in which the Earth is stationary, the dimensions of the planets' orbits are different from those of Brahe's and Ursus' systems, the orbits of Mars and the Sun are tangent, Saturn's orbit is tangent to the sphere of the fixed ones (in Ursus' and Brahe's systems there is some space) and the apparent orbits described by Mars and Jupiter around the Earth are also drawn (fig. 10).

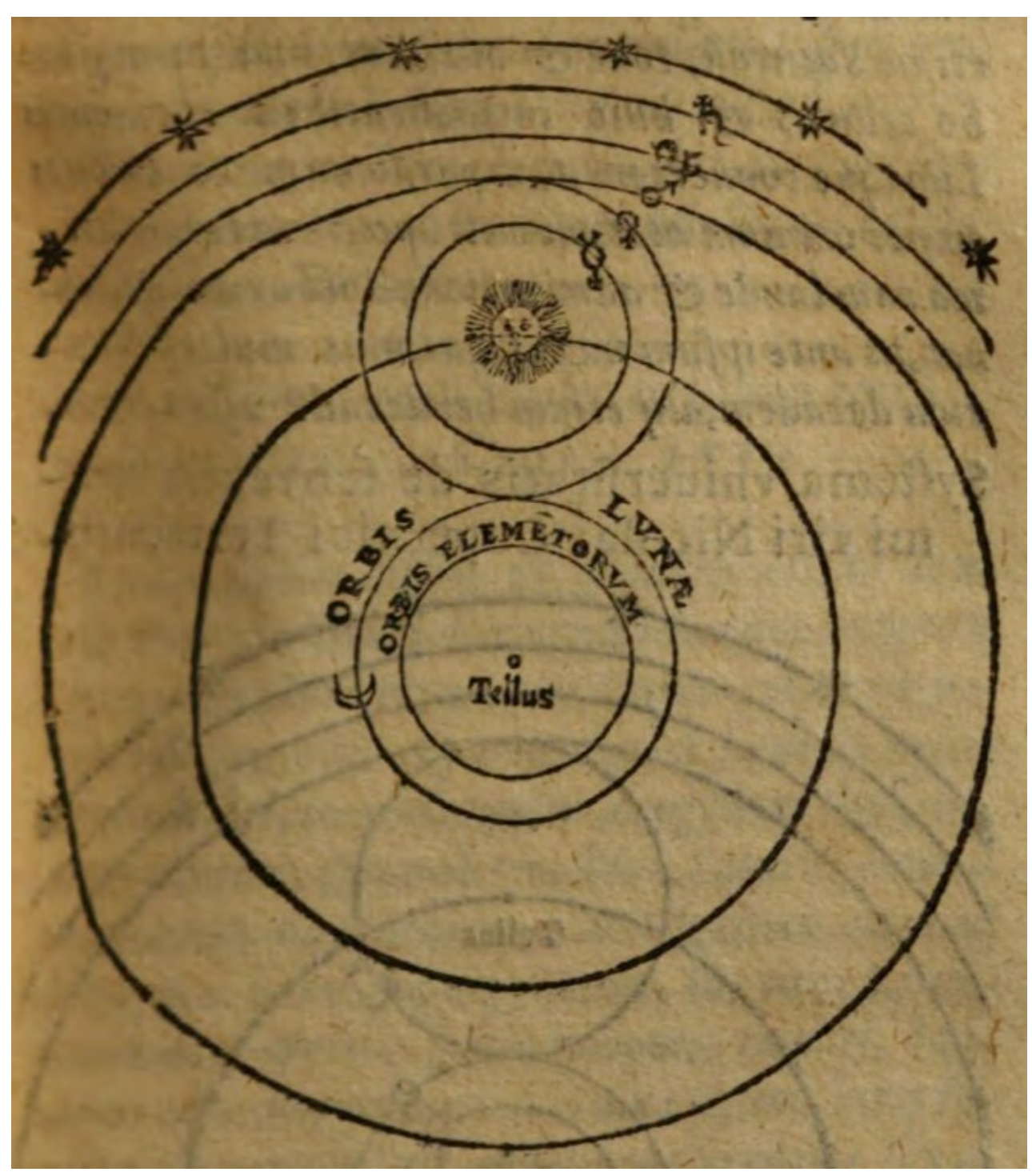

***Fig. 7.*** *The mixed system represented by Naboth (from* Primarum de coelo et terra institutionum quotidianarumque mundi revolutionum*).*

Several years later, Simon Mayr also claimed to have come up with the invention of a geo-heliocentric system in 1595-96,[68] but Mayr's unreliability, to put it mildly, does not speak in his favour.[69] Moreover, it is very unlikely that, however young he was at the time, he may never have heard of Tycho's system and the controversy over its priority (in fact, he says he learned of it the following year). He calls as his witnesses a large number of people, clergymen, councillors and teachers from the Heilbronn boarding school he attended, all of whom, alas, were already dead at the time of the claim, 1614.
In the decades that followed, there was some confusion in the literature as to the assignment of authorship of the geoheliocentric system, and it was only after 1621-22 that it basically began to be acknowledged to Brahe, being referred to it quite often as the "third world system".[70]

[60] Cf. O. Gingerich and R. S. Westman, *The Wittich connection*, "Transations of the American Philosophical Association", 78, 7, 1988, pp. 32 and 139.

[61] K. Ferguson, *The nobleman and his housedog*; I am quoting from the Italian edition: *L'uomo dal naso d'oro*, Longanesi, Milan, 2003, p. 138; Schofield, *op. cit.*, p. 52.

[62] O. Gingerich, *The book nobody read*; I quote from the Italian edition: *Alla ricerca del libro perduto*, Rizzoli, Milan, 2004, p. 156.

[63] *De mundi aetherei recentioribus phaenomenis*, Uraniborg, 1588, pp. 186 and 189.

[64] *Fundamentacum astronomicum*, Iobin, Strasbourg, 1588, ff. 37r-41r.

[65] Cf. Schofield, *op. cit.*, pp. 27-33.

[66] *Joannis Kepleri astronomi opera omnia* (ed. by Christian Frisch), Heider & Zimmer, Frankfurt and Erlangen, 1, 1858, pp. 229-230.

[67] *De opere dei creationis*, heirs Wechel, Marnius & Aubrius, Frankfurt, 1597. In the drawings illustrating the systems at the end of the work, captions have been mixed up between the systems of Tycho, Ursus and Copernicus.

[68] S. Mayr, *Mundus Jovialis*, Lauer, Nuremberg, 1615, f. C3r.

[69] Cf. G. Vanin, *On Simon Mayr's alleged discovery of Jupiter's satellites*, "Annals of Science", 81, 4, 2024, pp. 451–473; idem, *La polemica con Mayr sui satelliti di Giove*, "Atti del Convegno *Il Saggiatore di Galileo a 400 anni dalla sua pubblicazione*", Accademia Nazionale dei Lincei, 23-25 ottobre 2023, Bardi, Rome, 2024, pp. 311-324.

## 4. Inadequacy of the Tychonic system

The Tychonic system could at most be seen as a stepping stone towards full heliocentrism, and this is how Copernicus illustrates it in the first book of *De revolutionibus*, as a possible stage between Martian Capella's scheme (the one with Mercury and Venus orbiting the Sun) and the other Latin authors and his own.[71] At a time when Copernicus had already published his hypothesis, it could not but be considered a step backwards, a return to the past, a naive and useless compromise. That is, as an *ad hoc* theory (the kind that scientists gladly do without if they can, and, in this case, they could) contrived on purpose to have the best of both worlds, i.e. the immobility of the Earth and the concentric motion of the planets around the Sun, as a geometrically and physically poorly founded expedient. After Copernicus finally have the courage to declare that in the homocentric spheres, in epicycles and deferents lies hidden the orbit of the Earth, does Brahe return to the starting point, swapping the orbit of the Earth with that of the Sun again? It would be a bit as if, in the aftermath of the publication of the *On the Origin of Species*, someone had said that yes, evolution was acceptable, but that the evolution of man had occurred on a parallel track, not descending from the first common life-form, but from another one, completely different, and had developed quite independently. Or as if, in the aftermath of the proposition of the Big Bang theory, someone had said that yes, expansion was acceptable, but that our part of the universe had had an expansion subsequent to the initial one, and completely independent and unrelated to it.

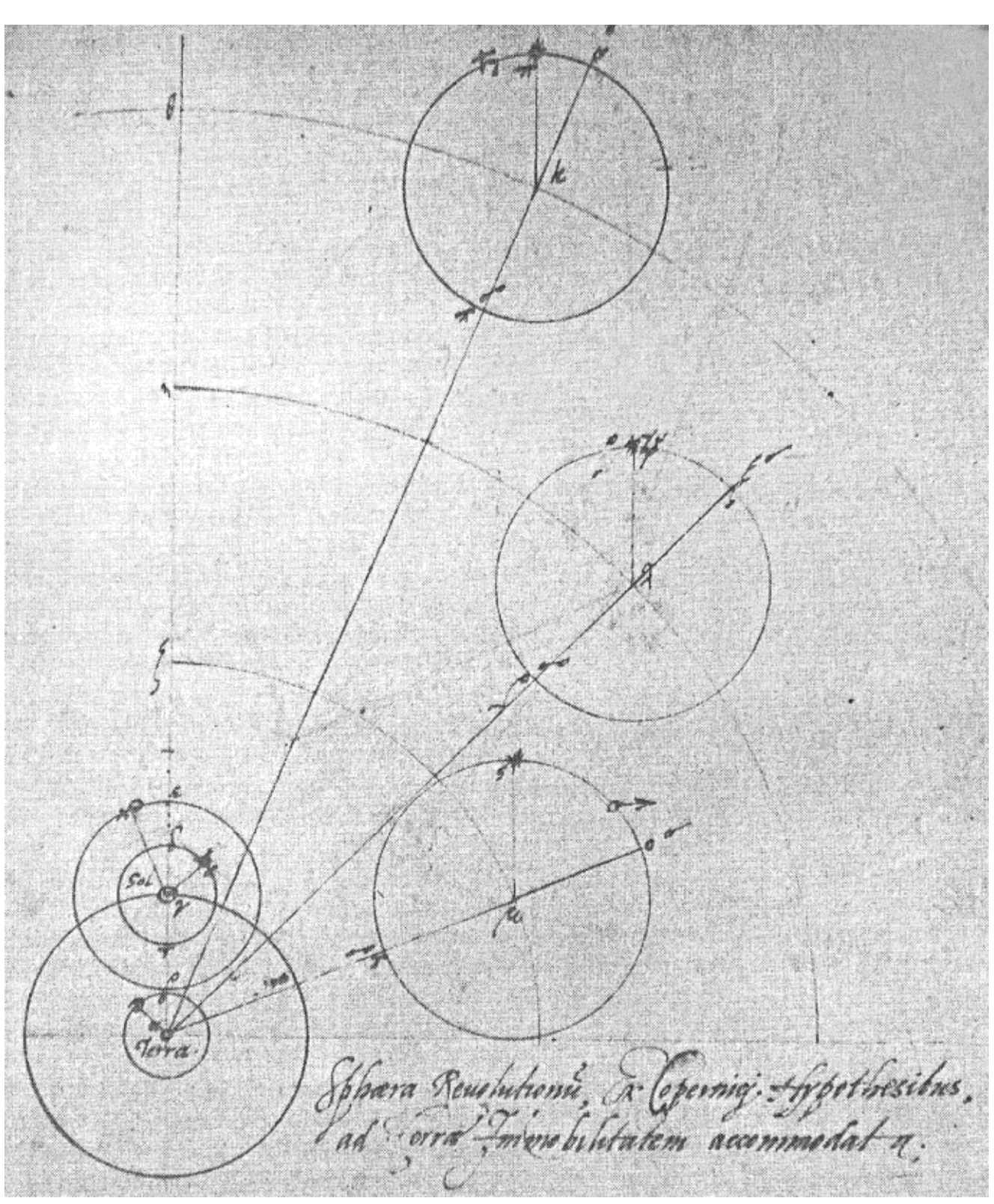

***Fig. 8.*** *Geoheliocentric scheme according to Capella drawn by Wittich in his copy of* De Revolutionibus *(Vatican Library, Coll. Ottoboniana 1902).*

Physically, the Tychonic system was only plausible if one continued to admit that a different type of physics applied in the heavens than that valid on Earth. In fact, the same perplexities that had steered Aristarchus towards heliocentrism could arise again: although for Tycho the Sun was a little closer than for Aristarchus (1150 Earth radii), our star had a volume 139 times greater than that of the Earth. Now, the belief that it was the large body that revolved around the small one was only possible if one continued to accept that celestial bodies were made of light and ethereal substance, as Aristotle said. But Tycho had already demonstrated, by measuring the parallax of the 1572 nova and proving that it belonged to the heavens, that they did not appear to be made of such ethereal and incorruptible substance at all. Moreover, when observing the 1577 comet, which seemed to transcend Aristotle's solid spheres, he had already demonstrated another fallacy of the Stagirite's physics. Why, then, still give him credit for the dichotomy between the lightness of the heavens and the heaviness of the Earth? Why continue to believe that, once the solid spheres that carried the planets had disappeared, the one to which the fixed stars were attached should continue to exist? Why continue to admit for it a motion contrary to all others, from east to west, instead of a motion of the Earth, from west to east, conforming to the others?[72] Why accept that what was moving was a sphere vastly larger than that of the Earth, and moreover moving at an immensely faster speed? The impression is that Tycho's thought was influenced not so much by physical arguments as by theological ones, since the Lutheran religion, much more so than the Catholic one, could not accept an allegorical interpretation of the Scriptures.

[70] Schofield, *op. cit.*, pp. 168-171.
[71] *De revolutionibus*, f. 8v.

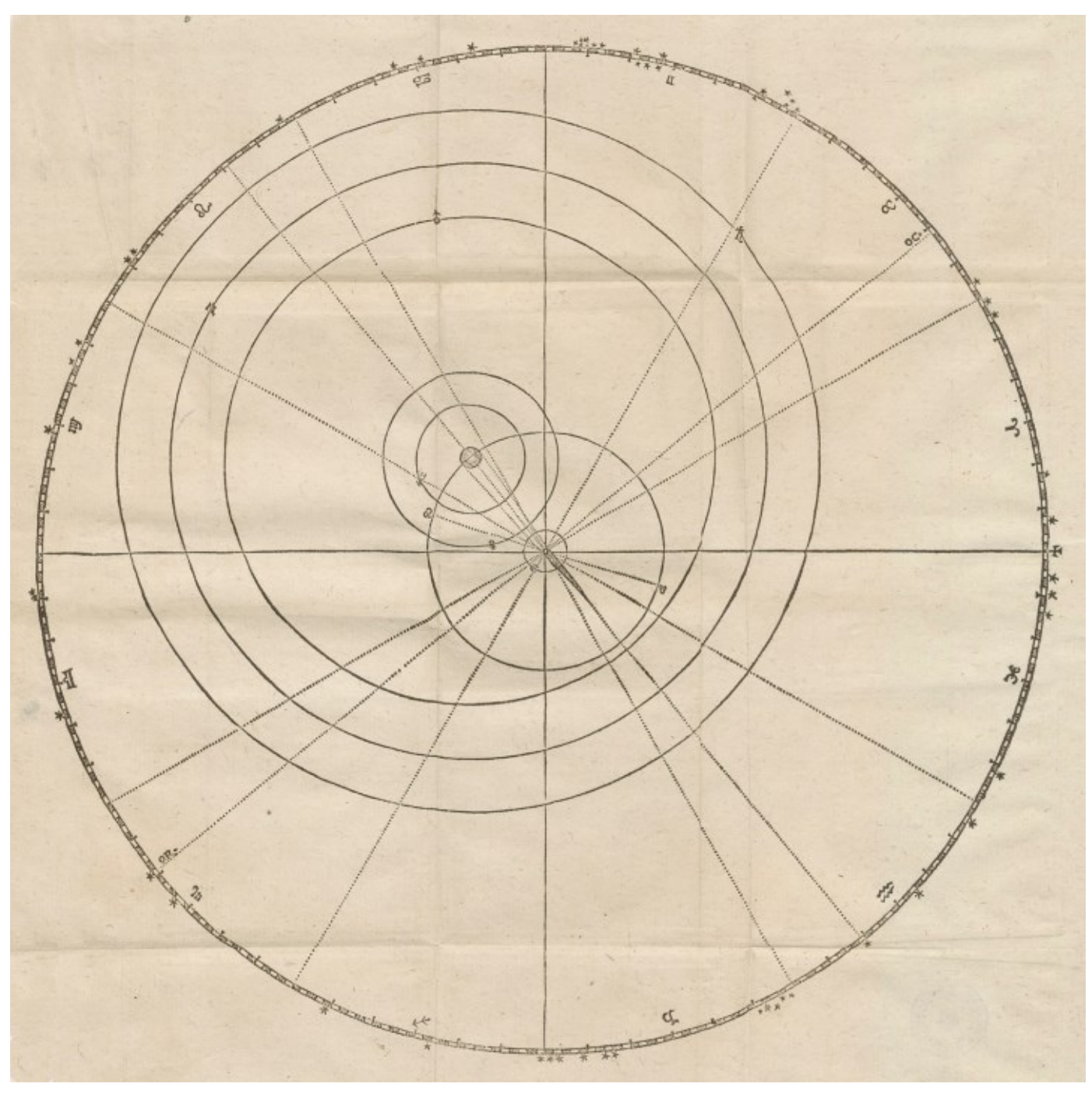

***Fig. 9.*** *The semi-Tychonic system of Ursus (from* Fundamentacum astronomicum*).*

As he was horrified by the errors in the planetary predictions of both the Ptolemaic and Copernican tables, Tycho's main aim was to found a new astronomy, based on observational reality, not on hypotheses. But in this concern, there is a gulf between his observations and his system of the world, in the sense that one struggles to understand how the former could have steered him towards the latter. Having spoken of the comet and the supernova, we have also already mentioned the contradiction inherent in the failure to detect the parallax of Mars. Contradictions also clearly emerge when considering the two main reasons, indeed the only reason, since the second one descends directly from the first, which drove Tycho to reject the Copernican system: the failure to detect stellar parallax, even with his highly perfected instruments, which allowed him, as mentioned above, to measure stellar parallax to an average accuracy of 1'. This value implied that the stars were at least 375 times the distance from Saturn, and Tycho was unable to explain what all that empty space was for. But this argument, in the light of his own observations, seems scarcely comprehensible: having demonstrated that there are no planetary spheres, and that therefore space is not filled by envelopes that end where the next one begins, and that planetary bodies are tiny compared to the trajectories they travel and the space that surrounds them, one sees no great difficulty in admitting spaces only an order of magnitude larger than that which separated, for example, Mercury from the Moon, 18 times the Earth-Moon distance. At the distance imagined by Brahe, as mentioned, it was necessary to admit that the stars had a colossal diameter, greater than the Earth's orbit itself. We have already seen that there were no valid objections in principle to this possibility, but it must also be said that it was based solely on the estimate of the truly enormous apparent diameters that Tycho assigned to the stars. This estimate is a truly gross error, for an observer as superbly trained as Tycho, and Galileo rightly chastised him: is it possible that he, and modern and past astronomers, had never observed Venus by day, or a star at twilight, to realise that their dimensions are much smaller (“without the hairs”) than in full darkness? That is to say:

[72] Cf. Galilei, *Dialogo*, p. 110.

> ...esteemed the true disc to be that which is seen in deep darkness, and not that which is seen in a luminous environment, because our lights, which appear large when seen at night from afar, and show their true, terminated, and small flame up close, could make them sufficiently cautious;...[73]

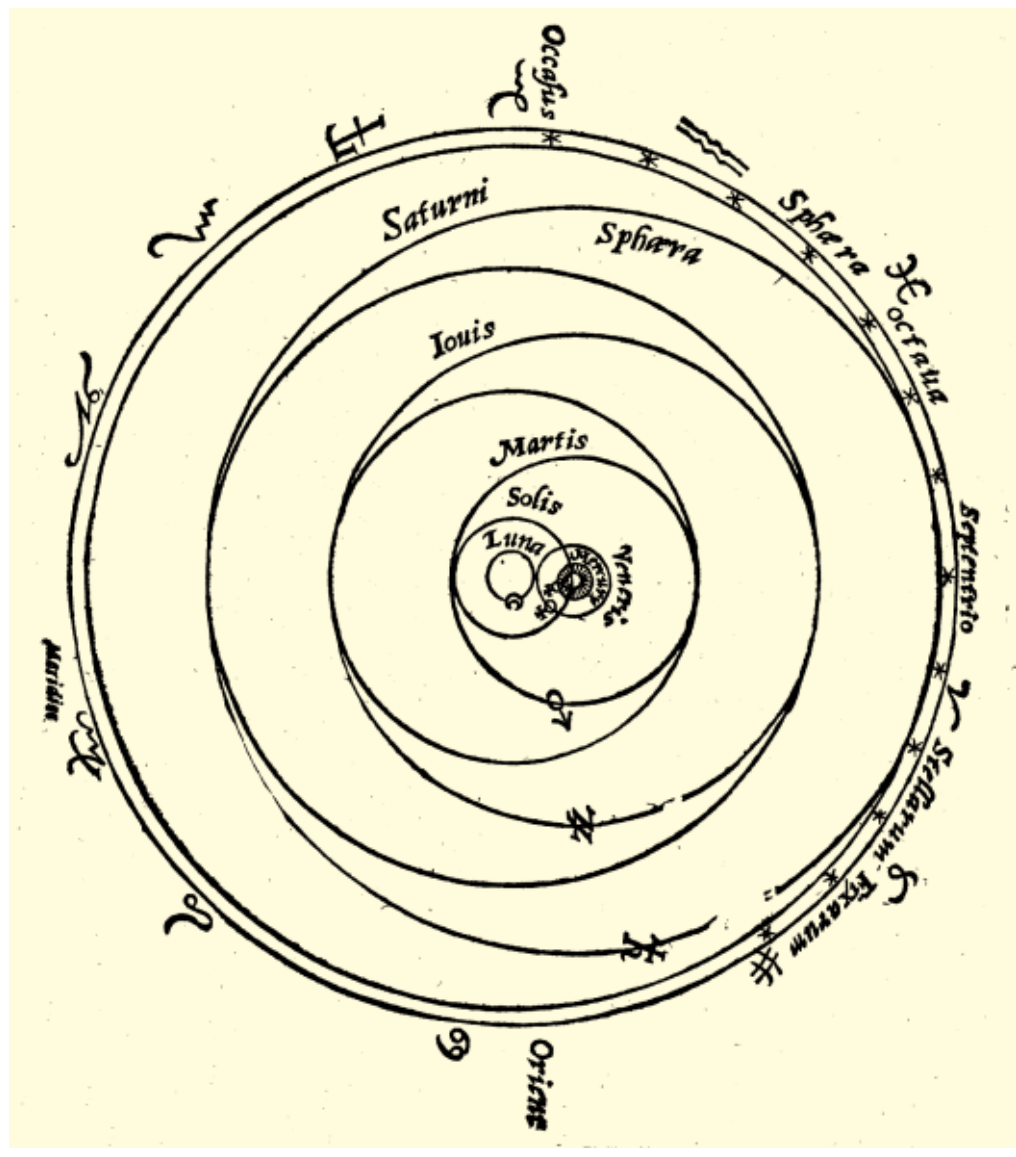

***Fig. 10.*** *Roeslin's cosmological system (from* De opere dei creationis*).*

Moreover, unlike the Ptolemaic and Copernican systems, as Galileo always notes,[74] the system that Tycho proposed remained only at a draft level and was not developed extensively in all its details, size of the deferents and epicycles, values of orbital eccentricities, number of epicycles for each planet, etc. Thus for many years prediction tables based on it were never constructed. Only Tycho’s pupil Christen Sørensen did so in 1622 and published the relevant tables.[75] It was not really Brahe's system, however, because Sørensen made the Earth turn on itself and, as mentioned, this was, physically and philosophically, quite another matter.
One also wonders who really adopted Brahe's system, apart from the Jesuit astronomers. Truth to tell, almost no one. No one in Italy, no one in England, in France only Jean-Baptiste Morin, in the Netherlands Erycius Puteanus[76] and Johannes Luyts,[77] in Denmark only Peter Bartholin,[78] no one in Spain, no one in Poland, David Fabricius in Germany,[79] no one in Bohemia, Austria or in the rest of the Holy Roman Empire.[80]

## 5. The Jesuits’ forced conversion

The Jesuits, as we know, were forced to convert to the geoheliocentric system because of the edicts of 1616 and 1632 by which the Church prevented them from professing Copernicanism. And it was not always the Tychonic system that they fell back on. For instance, Thomas Compton Carleton and Honoré Fabri leaned towards Capella's system, Joseph Zaragoza towards Riccioli's (see below), Gabriele Beati towards the semi-Tychonic one.[81]
And there are many indications that they were unwilling converts, starting from Clavius, who said about Tycho: “...it seems to me that it will never end, and that it confuses all astrology, since it wants Mars to be lower than the Sun”.[82] But also his disciples, first and foremost Grienberger, argued for Copernicanism and to them

---

[73] *Op. cit.*, p. 353.
[74] See footnote 2.
[75] *Astronomia Danica Christian S. Longomontani*, Coesius, Amsterdam, 1622, vol. 2.
[76] *Eryci Puteani de cometa anni 1618*, Butgenius, Cologne, 1619, p. 38.
[77] J. Luyts, *Astronomica institutio*, Halma, Utrecht, 1692, pp. 214-217.
[78] *Apologia pro observationibus, et hypotesibus astronomicis nobilissimi viri Dn. Tychonis Brahe Dani, a Petro Bartolino*, Moltke, Copenhagen, 1632.
[79] Dreyer, *op. cit.*, p. 369.
[80] See Schofield, *op. cit.*, pp. 289-308. Schofield mentions several more names, but a comparison with the original sources makes it possible to write off quite a few.
[81] *Op. cit.*, pp. 176-182.
[82] *Lettera di Cristoforo Clavio del 27 gennaio 1595 a Giovanni Antonio Magini*, in A. Favaro (ed.), *Carteggio inedito di Ticone Brahe, Giovanni Keplero, e di altri celebri astronomi e matematici dei secoli XVI. e XVII. con Giovanni Antonio Magini*, Zanichelli, Bologna, 1886, p. 215.

Tycho's system must have seemed a kind of philosophical monster.[83] It seems that this fact was also well known to contemporaries:

> ...because I mean that many Jesuits in secret are of the same opinion [the Copernican], though they keep silent:...[84]

> It is part of the new Creed, established by Pius IV in 1564, that no one should give assent to any interpretation of the Scriptures unless it is approved by the authority of the Fathers. And this is the reason why the Jesuits, who are the greatest advocates of those opinions, which seem new and ingenious, refrain from saying anything in defence of this [the Copernican system], but rather take every opportunity to rail against it.[85]

> And all the while the good Father Athanase [Kircher], whom we have seen pass by here in great haste, cannot refrain from warning us, in the presence of Father Ferrand, that Father Malapertius and Father Clavius himself do not disregard Copernicus' advice at all, but never depart from it, even though they have been urged and compelled to write for the common suppositions of Aristotle, which Father Scheiner himself follows only because he is forced to and does it out of a sense of duty.[86]

> ... and Father Scheiner's entire book [*Rosa ursina*] shows that they are not his [Galileo's] friends. But moreover, the observations in this book provide so much evidence aiming to take away from the Sun the movements attributed to it, that I could not believe that Father Scheiner does not share the opinion of Copernicus; ...[87]

## 6. Further difficulties

Even Tycho's biographer and editor of his *Opera Omnia*, Dreyer, treats the Brahe system condescendingly. He states among other things that: “The idea of the Tychonic system was so obvious a corollary to the Copernican system that it almost of necessity must have occurred independently to several people;...”.[88]
Even geometrically, as Kuhn pointed out,[89] it has its own inconsistencies, such as the fact that it appears skewed, with most of the planets shifted to one side, with the geometric centre of the cosmos not coinciding with the centre of most of the celestial motions, a far cry from the mathematical harmony of the Copernican universe.
Kepler had already pointed out that Brahe's system, by transferring the motion of the Earth to the Sun, unnecessarily multiplied the motions of the planets. Moreover, it was as complicated as and more so than Ptolemy's system, which concealed the motion of the Earth in the enormous epicycles of the planets, not least because it was necessary to compose the epicycles and deferents belonging to a single planet, but to combine the trajectories of the eccentricities of each planet with that of the Sun: the result, in the form of concatenated spirals, was even difficult to describe (fig. 11).[90] Another anomaly of the Tychonic system, according to Kepler, was the proportional jump in the orbital periods between Venus, 225 days, and Mars, 687 days: the Earth’s period, 365 days, intermediate between the two, was missing and it was transferred to the Sun on a trajectory that was, however, unnatural.[91]

---

[83] A. Fantoli, *Galileo per il copernicanesimo e per la Chiesa*, Specola Vaticana-Libreria Editrice Vaticana, Vatican City, 1993, pp. 119 and 213.
[84] *Lettera di Piero Dini a Galileo in Firenze del 16 maggio 1615*, in Galilei, *Opere*, 12, p. 181.
[85] J. Wilkins, *A discourse concerning a new planet*, John Maynard, Londra, 1640, p. 24.
[86] *Lettera di Niccolò Fabri di Peiresc a Pietro Gassendi in Digne*, 6-10 september 1634, in Galilei, *Opere*, 15, p. 254.
[87] *Lettera di Renato Descartes a Marino Mersenne*, February 1634, in Galilei, *Opere*, 16, p. 56. Campanella was also of the same opinion (*Apologia pro Galileo*, Kempffer, Frankfurt, 1622, p. 10) and Fabri de Peiresc reported that Scheiner had told him that he agreed with Galileo's opinion “on the worldly system” (*Lettera di Francesco Stelluti a [Galileo in Firenze]* of 10 January 1626, in Galilei, *Opere*, 13, p. 300.
[88] J. L. E. Dreyer, *History of the planetary systems from Thales to Kepler*, Cambridge at the University Press, 1906, p. 367.
[89] T. S. Kuhn, *The Copernican revolution*, Harvard University Press, Cambridge (USA) and London, 1957, p. 204.
[90] J. Keplerus, *Astronomia nova*, Prague, 1609, *Introductio* f. 3r; pp. 3-4.
[91] *Op. cit.*, *Introductio*, f. 3v.

Naturally, then, the Tychonic system did not square with the German astronomer's third law, enunciated in 1619. For if the Sun and Moon were satellites of the Earth, then the cubes of their semi-axes should have been proportional to the squares of their periods. This was not even remotely the case with the strong underestimation of the solar distance (which Brahe considered to be 20 times that of the Moon). Of course, in those days it might have been thought that the third law could only be valid for the Sun and its satellites, and not absolutely. But it had already been found to work for Jupiter's satellites,[92] discovered in 1610, and later also for Saturn's five, discovered between 1656 and 1684.[93] In short, Kepler's third law was strong evidence in support of the Copernican system and seemed to effectively falsify, as we would say today, the Tychonic system.[94]

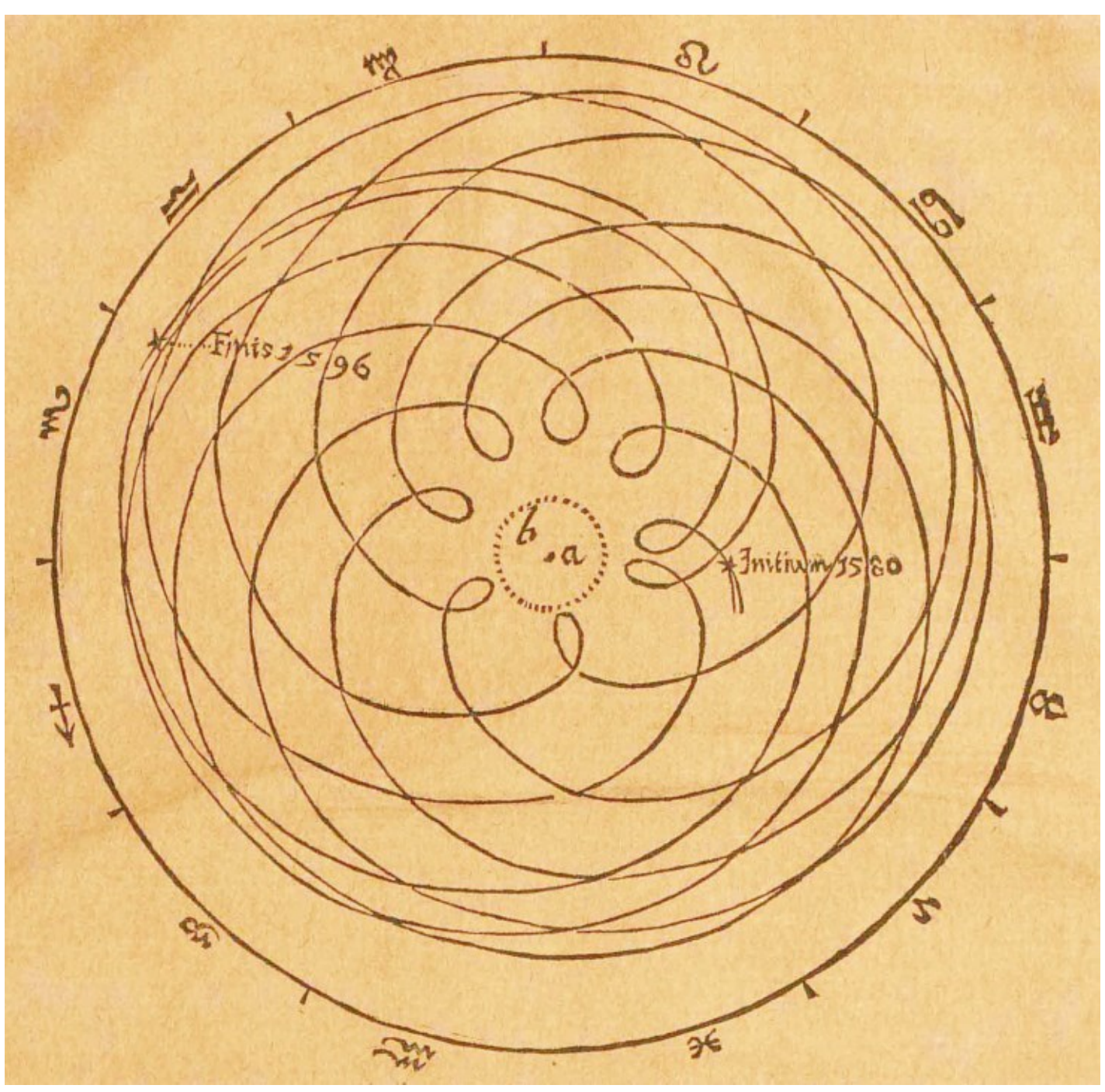

***Fig. 11.*** *Spiral motion limited to Mars alone, from 1580 to 1596 in the Ptolemaic and Tychonic systems, according to Kepler. The trajectories of the other planets are not shown here, nor those of the Sun around the Earth, nor their combinations (from* Astronomia nova*).*

The popularity of Brahe's system, which it experienced some forty years after its appearance, was almost exclusively due to the fact that, while Ptolemy had been disavowed by Galilean observations, Copernicanism had been forbidden by the Church, and thus there remained a scientific void to be filled, a deadlock to be overcome, after the rejection of the first two systems.

## 7. The Riccioli case

As for Riccioli, to continue to uphold the Tychonic system, albeit in a still different form (he had only Mercury, Venus and Mars revolve around the Sun, while the Sun, along with Jupiter and Saturn, revolved around the Earth, motionless on its own axis, fig. 12), may appear risky in the eyes of us moderns after the discoveries and interpretations of Galileo and the insights of Kepler, who had already begun to lay the foundations of a new physics 40 years earlier: the telescope seemed to prove that the planets and the Earth shared the same nature, and that there was therefore coherence between celestial and terrestrial bodies, along with Kepler's hypothesis that the planets moved by means of a force emanating from the Sun at the centre of the universe. And this does not mean diminishing the enormous scientific stature of the Jesuit, whose works constitute a veritable mine of information on the astronomy of his time.[95]

On the other hand, several historians have had strong doubts as to whether Riccioli was so convinced in his heart of the geo-heliocentric system. Delambre had already argued that he seems to be "a lawyer appointed

[92] Cf. Kepplero, *Epitome*, pp. 554-555.

[93] Cf. J. D. Cassini, *Découverte de la lumiere celeste qui paroist dans le zodiaque*, Imprimerie Royale, Paris, 1685, p. 22.

[94] If this did not happen, we know that it was because Kepler's ideas on planetary orbits did not find very fertile ground on the continent. It was only thanks to the much freer philosophical spirit of the "British" (Newton, Hooke, Halley, Wren) that those ideas brought about the accomplishment of the reform of astronomy long afterwards.

[95] And on which various studies and monographs have appeared in recent times; see e.g: I. Gambaro, *Giovan Battista Riccioli, infaticabile astronomo, e la censura romana*, "Giornale di astronomia", 42, 3, 2016, pp. 46-51 and F. Marcacci, *Un Gesuita contro tutti: astronomia e pensiero di Giovanni Battista Riccioli*", "Giornale di astronomia", 44, 3, 2018, pp. 11-20, with exhaustive bibliographical references to monographs and other studies.

to pleading a cause he scarcely knows, who only brings pitiful arguments because there are no others, and who knows himself that his cause is lost".[96] Whittaker is convinced that although Riccioli dedicated the Moon's brightest crater, with its system of rays, to Tycho, the choice of naming the three other sites with conspicuous ray structures after Aristarchus, Copernicus and Kepler implies that he thought the Copernican system was correct.[97] Monfasani stated that: "Riccioli expounded the Copernican system so extensively and generously that one would assume that, were it not for his obedience to the Holy See, he would have been a Copernican."[98]
In this regard, I would like to bring the following quote to the reader's attention:

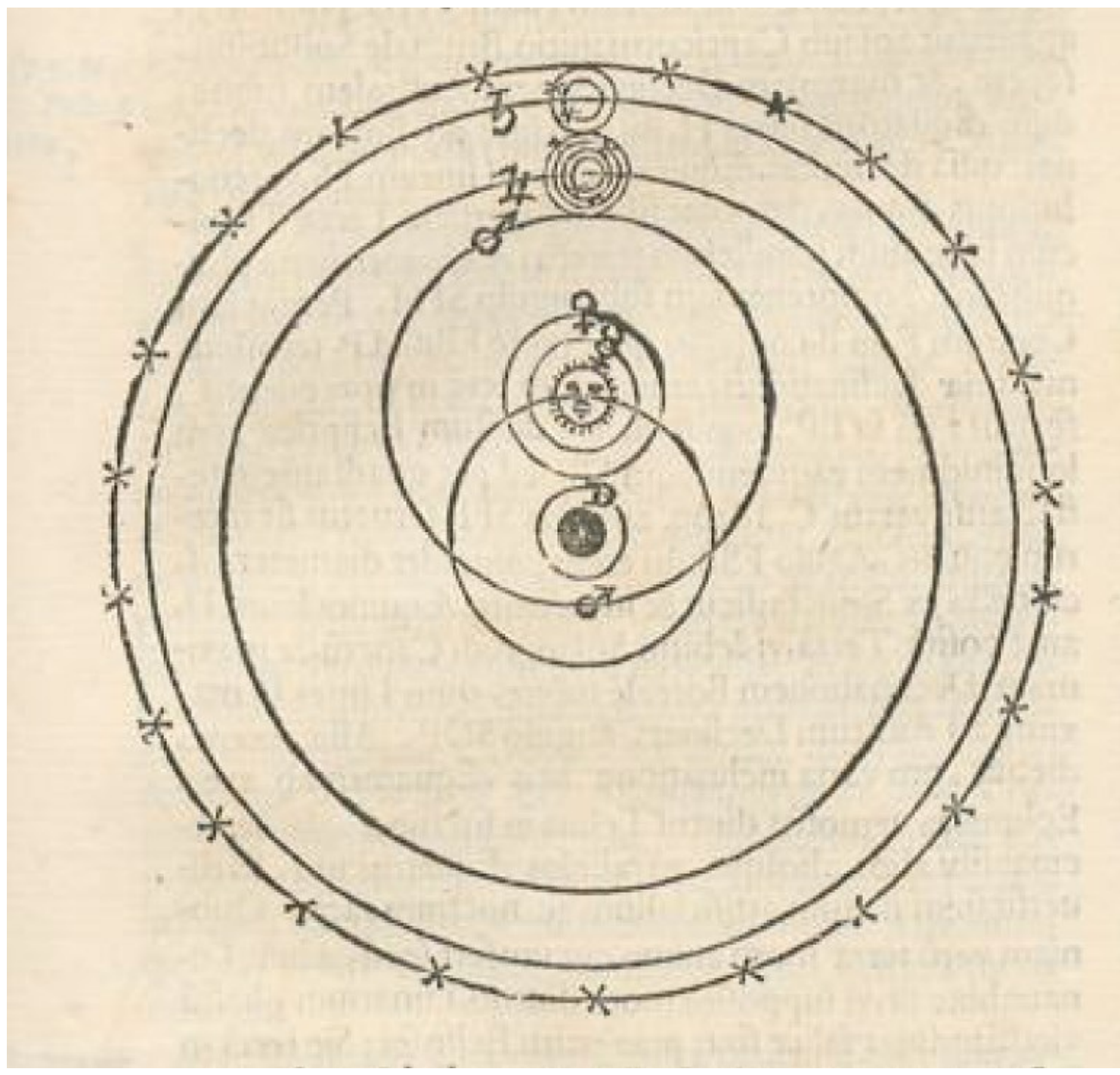

***Fig. 12.*** *Riccioli's cosmological system (from* Almagestum novum*).*

> Posterity has never sufficiently marvelled at the height of mind, the depth of soul, and the vigour of the genius of Copernicus, who, with the motion of a single small ball (how small the Earth is compared to the whole sky) tripled it, accomplishing what most astronomers before him could hardly have done without the mad devices of the spheres. For with the diurnal motion of the Earth he represented the first movable, and freed all other planets and fixed stars from that motion, which would otherwise have been accomplished by the vastest orbits and arrangements of circles, not without an apparent inconsistency with their own motion: and to the three eccentrics of the Sun, Venus, and Mercury, to the three epicycles of Saturn, Jupiter, and Mars, and to their second anomaly of motion, he substituted the single orbit and annual motion of the Earth; by which he also eliminated the apparent imperfection of the stations and retrogrades, and the oscillation of latitude and inclination with respect to the ecliptic. Finally, by means of a rotating oscillation of the earth's axis alone, he got rid of both the anomalies relating to the proper motion of the fixed stars and the variation of the obliquity of the ecliptic, i.e. he completely got rid of three huge spheres from the sky. And what before him so many Atlases could not, this Hercules dared to hold.[99]

I would not hesitate to call it the most effective promotional synthesis of the Copernican system ever written. Much more brilliant than what can be found in Copernicus himself (who was famously not a great promoter of his own system), and more effective even than what Galilei and Kepler themselves, taking many more pages, were able to put forward. One cannot disagree with what Delambre wrote, "that it would have been enough to remove the Jesuit's tunic to make him a zealous Copernican."[100]
A few decades later other Jesuits expressed flattering judgements on Copernicus' system, tempered more or less implicitly by the regret that they could not adopt it for scriptural reasons:[101]

---

[96] J. B. Delambre, *Histoire de l'astronomie moderne*, Courcier, Paris, 1821, 1, p. 681.
[97] E. A. Whitaker, *Mapping and naming the Moon*, Cambridge University Press, Cambridge, 1999, p. 65.
[98] J. Monfasani *Il rinascimento: scienza e religione*, in *Storia della Scienza*, 4, Istituto della Enciclopedia Italiana, Rome, 2001, p. 689.
[99] Riccioli, *op. cit.,* p. 309.
[100] Delambre, *op. cit*., p. 674. For a more detailed discussion on this point see Gambaro, *op. cit*.
[101] After all, Riccioli himself observed (did he have to observe?) at the end of the quote above: "If only he had been able to contain himself within the limits of this hypothesis."

> If it is used only as a hypothesis, this system, for the inequalities of motions, and appearances, and other properties we find in the Planets, explains them more easily and simply... It is really very ingenious because it saves and explains almost all appearances very easily and simply, without many epicycles and eccentrics...[102]

> No celestial phenomenon whatsoever lends itself to this hypothesis, indeed most of them are explained with fewer movements and circles than is usual;...[103]

> ...the authority of the Sacred Congregation of Cardinals, which in the sentence instituted on Galileo decreed the immobility of the Sun at the centre of the World and the annual Revolution of the Earth heretical and expressly contrary to Holy Scripture...[104]

> Copernicus explains these movements in the simplest manner of all, so much so that, if his hypothesis were not contrary to Scripture, it could be said to be absolutely divine. For, in order to achieve this, he explains these second periods from one conjunction to another, not by the true movements of the planets, but rather by an optical argument, so that the simple movement of the Earth, in the different parts of its orbit, produces all the modifications mentioned above, causing the observed configurations.[105]

## 8. Conclusions

In conclusion, the telescope observations of the first half of the 17th century did not favour the Tychonic system at all. The failure to measure stellar parallax even with the aid of the telescope could have perfectly logical and equally sensible explanations for the Earth's immobility. In the case of relative measurements, the stars could be physical pairs. Or they could be more or less at the same distance from us (their distance from each other not being such as to give rise to measurable parallaxes) and have different dimensions, without being physically linked. Or they could really be even further away from us, as Copernicus himself had already hypothesised:

> ... the sky is immense in comparison to the Earth, it appears to have the appearance of an infinite size, and based on the estimate of the senses the Earth is in comparison to the sky like a point in comparison to a body, like the finite in comparison to the infinite.[106]

Kepler for his part thought that the sphere of the fixed stars was 2000 times further away than Saturn.[107] Galileo himself must have suspected that the distances were indeed great, and in fact in his letter to Ingoli and in the *Dialogo* at one point he no longer speaks of double stars and suggests another system to arrive at the parallax measurement.
In the case of absolute measurements, e.g. in relation to the zenith, let me point out that Galilean telescopes, by their nature, could not be equipped with measuring devices in the focal plane; with Kepler's telescopes, Gascoigne's invention of the micrometer remained unknown before 1638; other, more perfected types of micrometer were developed by Huygens, Malvasia, Auzout, Wren, Hooke between 1659 and 1666, and not before the same period was the quadrant and sextant applied to a telescope.[108] It was thus impossible, for a long time, for this type of measurement to do truly better than Tycho Brahe's limit (1'). The first to succeed were the astronomers engaged in the campaign to measure the parallax of Mars (some of whom, as mentioned, also anticipated the discovery of the aberration of light), around 1670. Of course, if they were that far apart, and if the brightest stars were about 5” in diameter, as telescopic observation suggested, they must have been rather large, but not so dramatically: for a parallax of 30” the value is 0.08 times the orbit of the Earth. Moreover, Kepler already thought that the stars were mere points of light, without an apparent extension.[109] Jeremiah Horrocks demonstrated this experimentally, observing the instantaneous disappearance of the Pleiades stars during occultation by the Moon in 1637.[110] And I am not quite convinced that Galileo thought so too, given that in *Sidereus Nuncius* he expresses himself in this sense, that the estimates quoted in the Letter to Ingoli are upper limits, that those in the *Dialogo* are those suggested by the experiment being carried out, not by observation at the telescope, that those made at the telescope are reported in private, never published, writings.

[102] G. Beati, *Sphaera triplex artificialis, elementaris, ac* caelestis, Varesius, Rome, 1662, pp. 123-124.
[103] A. Tacquet. *Opera mathematica*, Meursius, Antwerp, 1669, p. 323.
[104] *Op. cit*., p. 331.
[105] C. F. Milliet de Chales, *Cursus seu mundus mathematicus*, Anissoniana, Lyon, 1674, 3, p. 287.
[106] *De revolutionibus*, f. 4*v*.
[107] Kepplero, *Epitome*, p. 497.
[108] R. Grant, *op. cit.*, pp. 449-456.

On the other hand, in Riccioli's time, a reasonable answer to the doubts raised by the anti-Copernicans had already come from Hortensius,[111] who had measured stellar diameters smaller than Riccioli's, from 2” to 10”, with his telescope. Assuming a parallax of no more than 30”, the stellar diameters ranged from 0.034 to 0.17 times the Earth's orbit, which seemed more than acceptable values.[112]

Ultimately, I believe that Galileo was right to exclude the Tychonic system from those worthy of scientific comparison on an equal footing, and in counting the Dane simply among the vast host of those who opposed the mobility of the Earth. I hope I have shown, even in this brief examination, that Brahe's system did not deserve all the consideration it earned, either at the time of its publication or in Galileo's time and after his death, because it could not constitute a truly credible alternative.

Historians generally believe that Galileo's evidence was not true evidence in favour of the Copernican system. If, however, as I have tried to show, the Tychonic system was out of the picture, the evidence Galileo collected at the telescope: on the one hand the “terrestrial” aspect of the Moon, the “corrupted” aspect of the Sun, the rotation of the Sun itself, the ashen light, the view of the planets as perceivable spheres, demonstrated the non-dichotomy between the sub- and superlunar worlds and thus the inconsistency of the Aristotelian view. On the other hand, the discovery of Jupiter's satellites, the way the sunspots moved, the phases of Venus, effectively falsified the Ptolemaic system. That is, they were just as much evidence in favour of the Copernican system. They certainly did not constitute the so-called “absolute proof” of the Earth's motion which many required, the so-called *experimentum crucis*,[113] according to a pre-Galilean scientific logic, inspired by pure empiricism of Baconian matrix, discriminating between Copernicus and Tycho. Still, Galileus’ various evidence was strong, consistent with his scientific principles and based on the value of the hypothesis that guides the collection of experimental data, both subjected to rigorous and rational analysis. So much so that when the “absolute proof” of the Earth's motion around the Sun, the aberration of light, was finally confirmed in 1729, there had not been a single follower of Tycho around for some time, if ever there had been any, of those truly free to choose on a philosophical and theological level.

But that is another story, and I will tell it in a future article.

[109] Kepplero, *Epitome*, pp. 497-498.

[110] A. B. Whatton, *Memoir of the life and labors of the rev. Jeremiah Horrox*, Wertheim, Macintosh & Hunt, London, 1859, p. 199.

[111] As also reported by Graney (*JHA*, 2010, p. 462), who nevertheless gives it much less attention than Riccioli.

[112] *Martini Hortensi delfensis dissertatio de Mercurio in Sole viso et Venere invisa*, Commelinus, Leiden, 1633, pp. 60-62.

[113] Concept referred to by Siebert, *op. cit*., pp. 251-252.